\pdfoutput=1
\documentclass[preprint,nofootinbib,tightenlines,superscriptaddress]{revtex4}
\usepackage{amsmath}
\usepackage{amsfonts}
\usepackage{amssymb}
\usepackage{graphicx, rotating}
\usepackage{epstopdf}
\usepackage{epsfig}
\usepackage{latexsym}
\usepackage{multirow}
\usepackage{color}
\usepackage{slashed}
\usepackage{mathrsfs,amssymb}

\newcommand{\ba}{\begin{eqnarray}}
\newcommand{\ea}{\end{eqnarray}}
\newcommand{\no}{\nonumber}
\newcommand{\be}{\begin{equation}}
\newcommand{\ee}{\end{equation}}

\usepackage[colorlinks]{hyperref}
\hypersetup{
    citecolor={blue}
}

\begin{document}

%\title{{\bf Neutrino Oscillations as unique probes \\
%of Secluded Dark Sectors}}

%\title{{\bf Neutrino Oscillations in Dark Backgrounds }}

\vspace{5cm}
%\title{\texorpdfstring{\PRE{\vspace*{0.5cm}}}{}  Neutrino Oscillations in Dark Backgrounds\texorpdfstring{\vspace*{0.5cm}}{}}

\title{ Neutrino Oscillations in Dark Backgrounds\texorpdfstring{\vspace*{0.5cm}}{}}

%Neutrino Oscillations and Dark Sectors

%\author{
%Francesco Capozzi$^{a,b}$\footnote{capozzi@mpp.mpg.de}, Ian M. Shoemaker$^c$\footnote{ian.shoemaker@usd.edu}, and Luca Vecchi$^{d}$\footnote{luca.vecchi@epfl.ch} 
%\\
%$^a${\small\emph{Center for Cosmology and AstroParticle Physics (CCAPP), Ohio State University, Columbus, OH 43210}}\\
%$^b${\small\emph{Max-Planck-Institut fur Physik (Werner-Heisenberg-Institut), Fohringer Ring 6, 80805 Munchen, Germany}}\\
%$^c${\small\emph{Department of Physics, University of South Dakota, Vermillion, SD 57069, USA}}\\
%$^d${\small\emph{Theoretical Particle Physics Laboratory, Institute of Physics, EPFL, Lausanne, Switzerland}}
%}

\author{Francesco Capozzi}
\email{capozzi@mpp.mpg.de}
\affiliation{Center for Cosmology and AstroParticle Physics (CCAPP), Ohio State University, Columbus, OH 43210}
\affiliation{Max-Planck-Institut fur Physik (Werner-Heisenberg-Institut), Fohringer Ring 6, 80805 Munchen, Germany}

\author{Ian M. Shoemaker}
\email{ian.shoemaker@usd.edu}
\affiliation{Department of Physics, University of South Dakota, Vermillion, SD 57069, USA}

\author{Luca Vecchi}
\email{luca.vecchi@epfl.ch}
\affiliation{Theoretical Particle Physics Laboratory, Institute of Physics, EPFL, Lausanne, Switzerland}

\date{\today}

\begin{abstract}

We examine scenarios in which a dark sector (dark matter, dark radiation, or dark energy) couples to the active neutrinos. For light and weakly-coupled exotic sectors we find that scalar, vector, or tensor dark backgrounds may appreciably impact neutrino propagation while remaining practically invisible to all other phenomenological probes. The dark medium may induce small departures from the Standard Model predictions or even offer an alternative explanation of neutrino oscillations. While the propagation of neutrinos is affected in all experiments, atmospheric data currently represent the most promising probe of the new physics scale. We quantify the future sensitivity of the ORCA detector of KM3NeT and the IceCube experiment and find that all exotic effects can be constrained at the level of a few percent of the Earth matter potential, with couplings mediating $\mu$-neutrino transitions being most constrained. {Long baseline experiments like DUNE may provide additional complementary information on the scale of the dark sector.}

\end{abstract}
\maketitle

\newpage

\section{Introduction}

There are two indisputable signs of physics beyond the Standard Model (SM). The necessity to accommodate neutrino oscillation data points to either a new mass scale, say the seesaw scale, or new light degrees of freedom, i.e. right-handed neutrinos. The second reason why the SM is incomplete is the experimental evidence of a dark world, which we commonly refer to as dark matter and dark energy, that has been found to permeate our galaxy and more generally the whole visible Universe.

For what we currently know, it may well be that the above two puzzles have a common physical origin. This exciting possibility may be motivated by the apparent similarity in magnitude of the two effects, which both appear to be of order $m_\nu\sim(\rho_{\rm DE})^{1/4}$. %{\color{blue}Perhaps a second, indirect argument is that the existence of exotic couplings between dark matter and neutrinos were identified as simple ways to alleviate some tension between observations and numerical simulations of primordial structure formation.} 
Next, there is a question of opportunity. The laws of Nature tell us there are only a handful of ways in which the dark sector can couple to our world without spoiling the tremendous success of the SM. One such option is to make use of the {\emph{neutrino portal}},~\footnote{By neutrino portal we mean a coupling to the gauge-singlet combination $H\ell$, with $H$ the SM Higgs and $\ell$ the lepton doublets. Together with the other two portals --- $|H|^2$ (Higgs portal) and the hyper-charge field strength $B_{\mu\nu}$ (kinetic mixing portal) --- they represent the SM-singlets with lowest dimensionality. Dark sectors with a neutral fermion, scalars, or $U(1)$ vector bosons, respectively, naturally couple to these operators.} that generally results in the introduction of exotic couplings between the active neutrinos and the dark world.

%The only way to introduce this effect consistently with SM gauge invariance is via the so-called {\emph{neutrino portal}}. In fact, the new physics associated with the dark world might couple to us via the neutrino portal. 

The existence of exotic couplings between active neutrinos and dark matter, dark energy or dark radiation, is expected to have {\emph{indirect}} astrophysical as well as cosmological consequences and, if we are lucky enough, even {\emph{direct}} signatures at colliders and dark matter experiments. Unfortunately, however, the absence of any clear indication as to where the new physics is hiding is suggestive that the latter is in fact too heavy to be directly accessible or possibly light but too weakly coupled to the SM. If this is indeed the case, then, is there any hope we will ever be able to directly access it? In this paper we suggest to use neutrino oscillations as a complementary probe of light, weakly-coupled dark scenarios coupled to us via the neutrino portal.

The coherent propagation of neutrinos in ordinary matter can amplify the otherwise small effects of the SM neutrino couplings to quarks and electrons~\cite{Wolfenstein:1977ue}\cite{Mikheev:1986gs}. In a similar way, neutrinos propagating in the Cosmo may be affected by hypothetical tiny couplings to the dark medium that surrounds us. What makes neutrino oscillation experiments unique compared to other probes of dark sectors, is their ability to exploit this coherent enhancement to probe secluded sectors that are extremely weakly-coupled and otherwise invisible~\cite{Capozzi:2017auw}. This feature of oscillation experiments becomes immediately obvious once we recall that the strength of the neutrino interactions to ordinary matter at low momentum transfers is controlled by the Fermi scale $G_F\sim g_W^2/m_W^2$, and realize that the latter quantity is actually a measure of the electroweak mass scale, and not of the electroweak coupling. In a world with a weak coupling $g_W$ smaller than ours by many orders of magnitude, the Fermi scale would still look the same to a physicist studying neutrino physics at momentum transfers $q^2\ll m_W^2$. Neutrino oscillation experiments take advantage of the forward scattering with the medium, in which $q^2=0$ and the $W$ boson mass only enters via $G_F$. The same argument generalizes to scenarios of physics beyond the SM: given a dark energy density $\rho_{\rm D}$, associated to dark matter or dark radiation or dark energy, neutrino forward scattering through this medium is sensitive to the combination $G_{\rm D}\rho_{\rm D}$, with $G_{\rm D}\sim g_{\rm D}^2/m_{\rm D}^2$ a ``dark Fermi constant" (we will see this in explicit models below). Secluded sectors with tiny couplings $g_{\rm D}\ll1$, that are not visible to experiments looking for processes that require sizable momentum transferred, can thus be probed in oscillation experiments if the mass scale characterizing the latter, $m_{\rm D}$, is small enough to compensate for the tiny density $\rho_{\rm D}\lesssim0.4$ GeV$/$cm$^3$. Small dark scales $m_{\rm D}$ appear in fact very reasonable, given that our current knowledge suggests that dark matter can be as light as $10^{-23}$ eV whereas the dark energy might be associated to fields with masses less than $H_0\sim10^{-33}$ eV.

In this paper we look at the signatures of neutrino propagation in a dark medium. In Section~\ref{sec:1} we describe how the ``effective Hamiltonian" relevant to neutrino oscillations is affected by interactions with general dark backgrounds. The outcome of this analysis, eq. (\ref{effH}), agrees with that obtained by earlier references, but we believe our derivation can provide useful to the reader intended to generalize the result including higher derivative interactions and other subleading effects. The exotic interactions are assumed to induce small corrections to the oscillation pattern characterizing the standard paradigm with three massive neutrinos. Yet, we will see in Section~\ref{sec:ether} that a dark ether with certain key properties might actually mimic the effect of neutrino masses and thus represent a {\emph{leading order effect}}. In such scenarios, neutrino oscillations themselves, as well as the relation $m_\nu\sim(\rho_{\rm DE})^{1/4}$, may be interpreted as the very first signature of the exotic interactions. 

Among all oscillation experiments, in Section~\ref{sec:4} we identify atmospheric neutrinos as key probes of these scenarios and estimate the experimental reach of two experiments. We argue that atmospheric data are, and will be in the foreseen future, necessary to constrain the relevant quantity $G_{\rm D}\rho_{\rm D}$. %Higher energy probes (astrophysical and GZK neutrinos) will become relevant (and eventually ... atmospheric) only after the sensitivity of atmospheric data has been improved enough. 
We show how the dark backgrounds can be generated in realistic models in Section~\ref{sec:models}. After a brief review of the scenario proposed in~\cite{Capozzi:2017auw}, we analyze an alternative vector model and the possibility of a dark radiation background, or dark CMB. This section demonstrates that oscillation experiments are unique probes of weakly-coupled dark sectors with masses of order keV or smaller. 

Earlier work has constrained interactions between neutrinos and the dark world, see e.g. ~\cite{Capozzi:2017auw, Fardon:2003eh,Kaplan:2004dq,Gu:2005eq,Ando:2009ts,Dvali:2011mn,Ciuffoli:2011ji,Berlin:2016woy,Capozzi:2017auw,Klop:2017dim,Brdar:2017kbt,Krnjaic:2017zlz,Davoudiasl:2018hjw}. {However, our paper is unique in recognizing the pivotal role of neutrino oscillation experiments in probing light and weakly-coupled dark sectors,} analyzing the effect of dipole interactions, and emphasizing the key role of atmospheric data. In our own~\cite{Capozzi:2017auw} we discussed a dark matter scenario capable of modifying solely solar data, but to achieve that feature some dark matter had to be captured in the sun, and this required additional interactions between dark and ordinary matter. Such additional interactions are not generic and not assumed here: in the present work no other interactions between the two sectors exist beyond those mediated by the neutrino portal. %A related work is by Kostelecky and c. on possible Lorentz and CPT violation. 

\section{General backgrounds}
\label{sec:1}

Let us assume that, similarly to interactions with ordinary matter, the couplings of neutrinos to the exotic world are small enough that one can safely ignore the back-reaction of neutrinos on the dark medium. We will quantify this statement when we discuss explicit scenarios in Section~\ref{sec:models}. Under this hypothesis, we can describe the dark medium as a non-dynamical background.

The most general quadratic Lagrangian for a set of Weyl fermions $N$ interacting with an external background can be written as
\ba\label{L}
{\cal L}&=&N_\alpha^\dagger\bar\sigma^\mu[i\partial_\mu\delta_{\alpha\beta}+(V_\mu)_{\alpha\beta}]N_\beta+\left\{\frac{1}{2}N_\alpha^\dagger\left[m_{\alpha\beta}+(F_{\mu\nu})_{\alpha\beta}\bar\sigma^{\mu\nu}\right]i\sigma_2 N_\beta^*+{\rm hc}\right\}\\\no
&=&{\cal L}_0+\delta{\cal L}\\\no
\delta{\cal L}&=&N^\dagger\bar\sigma^\mu V_\mu N+\left\{\frac{1}{2}N^\dagger F_{\mu\nu}\bar\sigma^{\mu\nu}i\sigma_2 N^*+{\rm hc}\right\},
\ea
with $\bar \sigma^\mu=(1,-\sigma^i)$, $\sigma^\mu=(1,+\sigma^i)$, $\bar\sigma^{\mu\nu}\equiv\bar\sigma^\mu\sigma^\nu-\bar\sigma^\nu\sigma^\mu$ (the latter differs from other conventions by a factor of $i/2$) and $\alpha,\beta,\cdots=e,\mu,\tau,\cdots$ are flavor indices.~\footnote{To avoid confusion, let us recall that Dirac neutrinos can be written as pairs of Majorana neutrinos (active and sterile), so our Lagrangian (\ref{L}) and the results of this section are completely general and apply irrespective of whether the total lepton number is conserved or not. The numerical analysis in Section~\ref{sec:4} will however assume Majorana neutrinos for definiteness.  
} Finally, $m,V,F$ are assumed to be generated by exotic couplings to the dark world. In Section~\ref{sec:models} we will see how this might happen in explicit models.

With respect to the flavor indices $\alpha,\beta=e,\mu,\tau,\cdots$, the mass is a complex symmetric matrix, the vector background is hermitian, and the tensor is a complex anti-symmetric matrix:
\ba\label{symmetry}
m_{\alpha\beta}=m_{\beta\alpha},~~~~~~~(V^\mu_{\alpha\beta})^*=V^\mu_{\beta\alpha},~~~~~~~F^{\mu\nu}_{\alpha\beta}=-F^{\mu\nu}_{\beta\alpha}.
\ea
As we will see more explicitly below, the vector background mediates particle/particle and anti-particle/anti-particle transitions, whereas the tensor induces a particle/anti-particle mixing among different flavors. The latter may in general involve sterile components.

In general $V,m,F$ may depend on space-time position and contain derivatives acting on $N$; for example, a local vector background would look like $V^\mu=V^\mu(x)+ V^{\mu\nu}(x)\partial_\nu+{\cal O}(\partial^2)$. We will first comment on space-time dependent backgrounds in Section~\ref{sec:ether} and then analyze in more detail constant backgrounds containing no derivatives starting from Section \ref{sec:constant}.

\subsection{Dark Backgrounds Faking Neutrino Masses?}
\label{sec:ether}

Our first task here is to show that under special conditions dark backgrounds as in (\ref{L}) can be the fundamental reason why neutrinos oscillate. This is easy to understand if we think of $m$, since a constant $m$ is indistinguishable from a mass.~\footnote{The possibility of relating $m_\nu$ to a dark scalar background ($m$ in (\ref{L})) was considered in earlier literature, see e.g.~\cite{Fardon:2003eh} and more recently in Refs.~\cite{Davoudiasl:2018hjw,DAmico:2018hgc}. Space-time dependent $m$ can be distinguished from ordinary masses as will be discussed in Section~\ref{sec:scalar}.} It is far less clear a priori that a vector and tensor background may accommodate neutrino oscillation data without the need of introducing neutrino masses, nor electroweak symmetry breaking. We will focus on $V^\mu$, but similar considerations apply to $F^{\mu\nu}$.%~\footnote{This follows from the dispersion relation (\ref{here}), with $(F_{\rm eff}^*)^{\mu\rho}(F_{\rm eff})_{\rho\nu}\propto\delta^i_j$ replacing (\ref{ass1}).}

Consider a background parametrized by a 4-vector $V^\mu$ that has a well-defined orientation within small domains of length size $\sim d$, but appears unpolarized at larger distances. More precisely, we assume that $V^\mu$ satisfies:
\ba\label{ass1}
\langle{V}^0\rangle,\langle{\bf V}\rangle\approx0,~~\langle{V}^i{V}^j\rangle\approx\frac{\delta^{ij}}{3}\langle{\bf V}^2\rangle,~~\langle{\bf V}^2\rangle\approx{\rm const},
\ea
where $\langle\cdots\rangle$ represents an average over distances $\gg d= {\cal O}(100)$ m. Indeed, experimental evidence of neutrino oscillations comes from local observations on scales of order hundreds of meters or larger. We therefore assume that Eqs. (\ref{ass1}) are satisfied within microscopic domains of size larger than such scales (and of course smaller than astronomical distances, i.e. the macroscopic scales on which the energy density of the dark world is expected to vary). Our picture has an obvious analogy with ferromagnets: the vector $\bf V(x)$ represents the magnetization density and depends on the space-time position, the microscopic distance $d$ is set by the ``magnetic domains", whereas the macroscopic scale is the size of the magnet itself. Our magnet has vanishing total magnetization at distances $\gg d$. 

Further assume the associated field has a flavor-violating coupling to the active neutrinos, as in eq. (\ref{L}). By hermiticity of $V^\mu_{\alpha\beta}$ we can diagonalize the latter via a unitary transformation (that, in the basis in which the charged leptons are diagonal, represents the PMNS matrix). The dispersion relation for a given neutrino ``$V^\mu$ eigenstate" --- where $V^\mu_{\alpha\beta}=\delta_{\alpha\beta}\hat V_\alpha^\mu$ --- reads $(p+\hat V)^2=0$ (see eq.(\ref{gen}) with $F=m=0$), that is
\ba\label{Vlead}
\langle E({\bf p})\rangle&=&\langle-\hat V^0+\sqrt{{\bf p}^2+{\bf\hat V}^2}\left[1+\frac{{\bf p}\cdot{\bf\hat V}}{{\bf p}^2+{\bf\hat V}^2}-\frac{1}{2}\left(\frac{{\bf p}\cdot{\bf\hat V}}{{\bf p}^2+{\bf\hat V}^2}\right)^2+\cdots\right]\rangle,
\\\no
&=&\left\{
\begin{matrix}
|{\bf p}|+\frac{1}{3|{\bf p}|}\langle {\bf\hat V}^2\rangle+\cdots~~~~~~~~~~~|{\bf p}|\gg|{\bf\hat V}|\\
\,|{\bf\hat V}|+\frac{1}{3|{\bf\hat V}|} {\bf p}^2+\cdots~~~~~~~~~~~~~|{\bf p}|\ll|{\bf\hat V}|,
\end{matrix}
\right.
\ea
where we neglected averages with 3 or more $V^i$s and used (\ref{ass1}). We see that for large as well as small 3 momenta ${\bf p}$ the effect of such a dark ether is practically indistinguishable from an effective neutrino mass $m_\nu\sim|{\bf V}|$. To an observer that cannot resolve the polarization of $V$, i.e. that cannot appreciate any departure from (\ref{ass1}), oscillations of relativistic neutrinos can be induced by $V^\mu$, and in particular proceed even in the absence of chiral symmetry breaking. This is somewhat analogous to the effect of a thermal mass at finite temperature. 

It is now tempting to relate $V^\mu$ to the dark energy and perhaps justify $m_\nu\sim(\rho_{\rm DE})^{1/4}$, though at present we see no special reason why such a $V^\mu$ should be associated to dark energy rather than dark matter or dark radiation. The main model-building challenge is to find a dark vector field with the properties in (\ref{ass1}). We will not attempt it here, however, and leave this task to future work. We simply note that, once an appropriate $V^\mu$ is found, obtaining the right coupling to neutrinos seems pretty straightfarward. For example, let us postulate there exists a scalar background $\phi$ whose gradient has inhomogenuities at small scales such that (\ref{ass1}) is satisfied. We can then couple it to the lepton doublets $\ell_\alpha$ via $c_{\alpha\beta}\ell^\dagger_\alpha\bar\sigma^\mu\ell_\beta\partial_\mu\phi/f$ and effectively obtain $V^\mu_{\alpha\beta}=c_{\alpha\beta}\partial^\mu\phi/f$, that can mimic neutrino oscillations if $\phi$ has spacial inhomogenuities of the right size. Corrections to the dispersion relation of the charged leptons are of order $<m_\nu/m_\ell\sim10^{-10}$ and therefore too small to be constrained by current data. Flavor-violating processes like $\mu\to e\gamma$ do not represent a serious threat because sufficiently suppressed if $f$ is large enough.

It is interesting to consider the typical signatures of this scenario. From (\ref{Vlead}) we see that the effective mass is different at high and low energies: in the relativistic limit $m_\nu^2=\frac{2}{3}{\bf V}^2$ whereas in the non-relativistic limit $m_\nu=\frac{3}{2}|{\bf V}|$. This latter aspect represents a characteristic signature that may be tested, for example, studying relic neutrinos or beta decay. Further tests could possibly come from extra-galactic neutrinos whose flavor ratio could be altered as they enter or exit the dark background.

In addition, there are other signatures one can consider. While the leading order effect described in (\ref{Vlead}) is practically indistinguishable from those of an actual neutrino mass, departures from the idealized limit (\ref{ass1}) are characterized by corrections $\langle-V^0+{\hat{\bf p}\cdot{\bf V}}+\cdots\rangle\neq0$ to the dispersion relation (\ref{Vlead}). These can be effectively parametrized by \emph{small and constant} $m,V,F$. Importantly, corrections to the standard paradigm of three massive neutrinos parametrized by small and constant $m,V,F$ are common to all our scenarios, whatever the UV origin of neutrino oscillations is. The rest of the paper is thus devoted to constrain those effects.

\subsection{Constant dark backgrounds}
\label{sec:constant}

In the remainder of the paper we will focus on backgrounds that are approximately constant within the length scales of interest, from a hundred meters to the earth diameter or longer. This approximation is justified because the distributions of dark matter and dark energy presumably vary on (extra-)galactic distances. And, as we have seen in Section \ref{sec:ether}, even in scenarios with inhomogenuities at shorter distances, the most significant departure from the standard oscillation framework is still parametrized by the effect of (\ref{L}) with approximately constant $m,V,F$.  

Furthermore, terms with additional derivatives acting on $N$ are usually subdominant and can be ignored. This is easy to understand in scenarios in which the interactions leading to (\ref{L}) are mediated by heavy fields with masses of order a scale $\Lambda$ that is much larger than the typical neutrino energy $E$. In those cases extra derivatives acting on $N$ would result in effects suppressed by powers of $E/\Lambda\ll1$ compared to those without derivatives. A similar conclusion generalizes to scenarios with light mediators. To be concrete, let us consider perturbative models --- in which the dark backgrounds originate from the vacuum of spin $0,1$ fields $\phi,A_\mu$. In these cases the higher order tensors necessary to build interactions with extra derivatives on $N$, as for instance $V^\mu\supset V^{\mu\nu}\partial_\nu$, must be obtained differentiating $\phi,A_\mu$; but these derivatives are typically suppressed under the hypothesis of approximate homogenuity of the dark medium. 

These simple considerations motivate us to neglect any space-time dependence, as well as possible derivative interactions, in $m,V,F$ and focus on the more relevant case 
\ba\label{hypo}
m,V_\mu,F_{\mu\nu}~=~{\rm constant~backgrounds}.
\ea
%The microscopic origin of the various terms in (\ref{L}) is model-dependent. We will discuss a few representative explicit realizations in Section~\ref{sec:models}. 
Approximately constant vector and tensor backgrounds as in (\ref{hypo}) signal a local violation of Lorentz invariance completely analogous to that implied by solids (made up of dark stuff, in our case). Also, CPT is locally broken by $V\neq0$. Therefore the present paper can also be seen as an analysis of the {\emph{leading}} effects of Lorentz and CPT violation. Our work has thus some overlap with~\cite{Kostelecky:2003cr}\cite{Kostelecky:2004hg} (for a summary of the current constraints the reader may refer to \cite{Kostelecky:2008ts}).~\footnote{Many papers on this topic also include the effect of higher-derivative couplings such as $V^{\mu\nu}\partial_\nu$. As argued above, however, these terms are expected to be subleading.} 

%Experimental bounds on Lorentz and CPT Violation have been analyzed in several papers (see e.g.~\cite{Katori:2006mz} for a first global analysis, \cite{Diaz:2016fqd} for a recent discussion of solar neutrinos and~\cite{Abe:2014wla} PLUS IceCube!!! for atmospheric neutrinos. . Diaz is the theorist currently working on this~\cite{Diaz:2009qk}. 

Let us now turn to a discussion of how neutrino oscillations are affected by the general Lagrangian (\ref{L}) in the interesting cases described by (\ref{hypo}).

\subsubsection{The ``Effective Hamiltonian"}

In all instances of physical relevance, neutrino oscillations can be understood as the measure of the probability of a transition between stationary states $|\nu\rangle$ with definite flavor, helicity, and energy traveling a spacial distance $L$, see~\cite{Lipkin:1995cb}\cite{Stodolsky:1998tc} (and \cite{Lipkin:1999nb} for a more explicit discussion). The amplitude for such processes may be derived by calculating $\langle\nu'|e^{iPL}|\nu\rangle$, with $P$ the operator associated to the 3-momentum along the flux and $L$ the distance between source and detector, i.e. the experimental baseline.

A careful derivation of $P$ is presented in Appendix~\ref{app:} in the limit of small constant $V,F$ and ultra-relativistic neutrinos~\footnote{Because $V,F$ are Lorentz tensors, the term ``small" is frame-dependent. Here we implicitly assume that $V,F$ are small with respect to an observer in the galactic or earth frames.}. One can re-state our result (\ref{Vint}) in a more familiar fashion. In the Schroedinger basis, and up to an unphysical phase, the propagation between source and detector may be described via $i\frac{d}{dx}|\nu(x)\rangle=\delta H|\nu(x)\rangle$, where $\delta H$ is what is commonly referred to as effective ``Hamiltonian" (though it is in fact an effective 3-momentum). Then, oscillations of properly normalized wave-packets with definite flavor and helicity are controlled by a $(2n_N)\times (2n_N)$ matrix, with $n_N=1,2,3,\cdots$ the number of Weyl neutrinos:
\ba\label{effH}
H_{\rm eff}\equiv\langle \nu'|\delta H|\nu\rangle=\left(
\begin{matrix}
U\frac{mm^*}{2E}U^\dagger+a(\hat{{\bf p}}) & b(\hat{{\bf p}}) \\ 
b^\dagger(\hat{{\bf p}}) & U^*\frac{m^*m}{2E}U^t-a^*(\hat{{\bf p}})
\end{matrix}
\right).
\ea
Here
\ba\label{ab}
&&a(\hat{{\bf p}})=-V_0+{\hat{\bf p}}\cdot {\bf V}\\\no
&&b(\hat{\bf p})=-4\sqrt{2}({\bf E}+i{\bf B})\cdot\vec\epsilon(\hat{\bf p}),
\ea
with $|{\bf p}|=E$ up to subleading corrections, $\hat{\bf p}={\bf p}/|{\bf p}|$, and $\vec\epsilon(\hat{\bf p})$ is defined in eq.(\ref{Vint}). The properties (\ref{symmetry}) imply that $a(\hat{{\bf p}})=a^\dagger(\hat{{\bf p}})$ and $b(\hat{{\bf p}})=-b^t(\hat{{\bf p}})$, where complex conjugation and transposition are in the flavor indices and both matrices are allowed to be complex. In (\ref{ab}) we introduced the dark ``electric" and ``magnetic" fields $F_{0i}=E_i$ and $F_{ij}=\epsilon_{ijk}B_k$ and used the relation ${\bf p}\wedge\vec\epsilon=-i|{\bf p}|\vec\epsilon$. The expressions (\ref{ab}) make it manifest that $a,b$ in general do depend on the neutrino direction. In particular, while an isotropic vector background is possible if ${\bf V}=0$, the tensor interaction necessarily breaks rotational invariance by defining a preferred direction.~\footnote{In the ultra-relativistic regime the dipole is {\emph{orthogonal}} to the direction of motion, $\vec\mu\propto\vec\epsilon$, and therefore to the helicity. This is perhaps unfamiliar because a non-relativistic particle has a dipole moment proportional to its spin, $\vec\mu\propto\vec s$. What the two regimes have in common, however, is that the dipole ceases to have physical effects when the external field is in the same direction as the particle's spin. It is only when the polarization has a component orthogonal to the field that we see a ``precession", that here corresponds to a transition between different helicities.}

Note that, at the leading order in the perturbations, eq. (\ref{effH}) coincides with the {\emph{Hamiltonian}} found using a quantum mechanical approach in~\cite{Kostelecky:2003cr}. This is because under the assumption of constant backgrounds the momentum operator coincides with the one in the absence of the perturbations ($P=P_0$ in Appendix~\ref{app:}), and so small perturbations $\delta H$ in the Hamiltonian effectively translate into a correction of the 3-momentum according to (\ref{Pgen}). In the presence of additional space-time derivatives or strongly varying backgrounds the two approaches would in general give different results, however. %In particular, if higher derivatives are included the simple relation $\delta H_{\rm int}(t)=-\int d{\bf x}~\delta{\cal L}[N_0,N^\dagger_0]$, that we used in the appendix, is no more valid.

It is possible to understand our result (\ref{effH}) without doing any actual computation. Indeed, it is important to appreciate that the necessity of having appropriate contractions of the flavor indices $\alpha,\beta$ to render $H_{\rm eff}$ covariant uniquely determines where the various exotic backgrounds appear in (\ref{effH}). In particular the flavor symmetry implies that $mm^*,V$ must be associated to particle/particle mixing and $m^*m,V^*$ to anti-particle/anti-particle mixing and must therefore appear in the diagonal elements of (\ref{effH}). On the other hand, the only place where $F$ can show up is in the off-diagonal term, thus inducing particle/anti-particle transitions. This can be interpreted as the consequence of a {\emph{flavor selection rule}} of (\ref{L}) that forces $a,b,mm^*$ to transform as $a\to U_\nu aU_\nu^\dagger$, $mm^*\to U_\nu mm^*U_\nu^\dagger$, and $b\to U_\nu bU_{\nu}^t$, where $U_\nu$ is an arbitrary rotation in neutrino flavor space. The need to properly contract the $U_\nu$ indices to form physical, re-phasing-invariant objects also selects the way $a,mm^*,b$ appear in the oscillation probabilities, as we will see below.

The numerical factors in front of the various entries of (\ref{effH}) can be confirmed inspecting the eigenvalues of the Hamiltonian operator, that themselves follow from the neutrino dispersion relation:
\ba\label{gen}
0={\rm det}\left[(\bar{\slashed{p}}+\bar{\slashed{V}})-(m+F_{\mu\nu}\bar\sigma^{\mu\nu}){({\slashed{p}}-\slashed{V})}^{-1}(m^*+F^*_{\mu\nu}\sigma^{\mu\nu})\right].
\ea
Here all entries are understood as matrices in flavor space. The problem of finding solutions $p^0=E({\bf p},m,V,F)$ to (\ref{gen}) is considerably simplified when working at leading order in the perturbations. Within this approximation the mixed terms can be neglected, as can be seen employing the flavor selection rules mentioned above. This semplification implies that $E=|{\bf p}|+\cdots$, where the dots contain three independent corrections proportional respectively to $m,V,F$. We can now derive the leading corrections by solving three distinct dispersion relations, where $m,V,F$ are switched on one at a time. The case $V=F=0$ is standard and will not be described here. To study the effect of $V$ we switch off $F$ and neglect mixed terms involving multiplication of $m$ and $V$. Eq. (\ref{gen}) is left invariant by the transformation $(m,V)\to(m^*,-V^*)$. It follows that the two positive energy states have, at the order of interest, energies $E_+=E({\bf p},m,V)$ and $E_-=E({\bf p},m^*,-V^*)$. Within the conventions in the Fourier transform adopted in (\ref{gen}), particles (helicity $-$) have $E_-=|{\bf p}|-p^\mu V_\mu/|{\bf p}|+\cdots$ and anti-particles (helicity $+$) have $E_+=|{\bf p}|+p^\mu V_\mu^*/|{\bf p}|+\cdots$ consistently with (\ref{effH}). The vector enters differently in the energy of particles and anti-particles because it is odd under CPT. Finally, to derive the dispersion relation for $F\neq0$ we set $m=V=0$ and observe that ${\rm det}\left[\bar{\slashed{p}}-F_{\mu\nu}\bar\sigma^{\mu\nu}{\slashed{p}}^{-1}F^*_{\rho\sigma}\sigma^{\rho\sigma}\right]=0$ can be diagonalized in flavor space. In that basis the determinant separates and, after some work, we find that for each flavor index $\alpha$ the dispersion equation reads
\ba\label{here}
0=(p^2)^2+\sum_\beta{\rm tr}[\slashed{p}\bar F_{\alpha\beta}\slashed{\bar p}F_{\alpha\beta}]+{\cal O}(F^4)=(p^2)^2-32\, p_\mu (F_{\rm eff}^*)^{\mu\rho}(F_{\rm eff})_{\rho\nu}p^\nu+{\cal O}(F^4),
\ea
where we introduced $(F_{\rm eff})^{\mu\nu}=(F_{\alpha\beta})^{\mu\nu}-\frac{i}{2}\epsilon^{\mu\nu\rho\sigma}(F_{\alpha\beta})_{\rho\sigma}$ and in the second step of~(\ref{here}) the sum over the flavor index $\beta$ is understood. For each flavor $\alpha$ there are two solutions $E-|{\bf p}|=\pm{\sqrt{8\, p_\mu (F_{\rm eff})^{\mu\rho}(F_{\rm eff}^*)_{\rho\nu}p^\nu}}/{|{\bf p}|}+{\cal O}(F^2)$, with $p^\mu=(|{\bf p}|,{\bf p})$. These coincide with those derived from (\ref{effH}).~\footnote{In the case of two flavors the relation $[4\sqrt{2}p^\mu F_{\mu\nu}\epsilon^\nu][4\sqrt{2}p^\mu F^*_{\mu\nu}(\epsilon^\nu)^*]=8\,p_\mu (F_{\rm eff})^{\mu\alpha}(F_{\rm eff}^*)_{\alpha\nu} p^\nu$ can be readily used to show that $E-|{\bf p}|=\pm|4\sqrt{2}p^\mu F_{\mu\nu}\epsilon^\nu|/|{\bf p}|$, consistently with (\ref{Vint}). The generalization to an arbitrary number $n_N$ of flavors follows. Hence we can conclude that (\ref{Vint}) correctly reproduces the leading energy shift in the general case as well.}

\section{Future constraints from Atmospheric Data}
\label{sec:4}

\subsection{Why atmospheric data}

The effective Hamiltonian (\ref{effH}) can be schematically written as the sum of the SM term $H=U\frac{\Delta m^2}{2E}U^\dagger+V_{\rm SM}$ and an exotic piece $\delta V_{\rm eff}$ proportional to $a,b$. It is useful to compare the potential of the main classes of neutrino oscillation data that can be used to set limits on $\delta V_{\rm eff}$: astrophysical, atmospheric, reactor, and solar.

Let us first assume that the exotic term leads to a small correction to the SM oscillation probability. This is equivalent to the statement that $\delta V_{\rm eff}$ is small compared to the SM vacuum Hamiltonian. Assuming $\delta V_{\rm eff}\ll H$ we have:
\ba
P(\alpha_\lambda\to\beta_\lambda)%&=&\left|(e^{-iHL})_{\beta\alpha}\right|^2+2\,{\rm Im}\left[(e^{-iHL})^*_{\beta\alpha}\left(\int_0^L dx\,e^{-iH(L-x)}\delta V_{\rm eff}e^{-iHx}\right)_{\beta\alpha}\right]+{\cal O}(\delta V_{\rm eff}^2)\\
\label{scalingP}
&=&P_{\rm SM}+{\cal O}(\delta V_{\rm eff}{\rm min}(L,L_{\rm osc}))~~~~~~~~~~~~~~~({\rm for}~~~\delta V_{\rm eff}\ll H)\\
P(\alpha_\lambda\to\beta_{\lambda'\neq\lambda})%&=&\left|\int_0^L dx\,e^{-iH(L-x)}\delta V_{\rm eff}e^{-iHx}\right|^2+{\cal O}(\delta V_{\rm eff}^3)\\
\label{scalingP'}
&=&{\cal O}(\delta V_{\rm eff}{\rm min}(L,L_{\rm osc}))^2~~~~~~~~~~~~~~~~~~~~~~\,({\rm for}~~~\delta V_{\rm eff}\ll H),
\ea
where $\lambda,\lambda'$ indicate the helicity. The oscillation length $L_{\rm osc}$ is controlled by the SM and is of course set by $\Delta m^2/E$. The scaling laws in (\ref{scalingP}) (\ref{scalingP'}) can be shown in complete generality, and are of course reproduced by the examples in Section~\ref{sec:examples}. Note that at short baselines only the off-diagonal entries in $\delta V_{\rm eff}$ can be constrained by $P(\alpha_\lambda\to\beta_\lambda\neq\alpha_\lambda)$ measurements, whereas long baseline experiments in principle have access to all elements because of the flavor transitions occurring in the SM. 

The new physics term $\delta V_{\rm eff}$ does depend on the neutrino direction but is otherwise approximately constant in energy, see (\ref{effH}). Hence, the relevance of the exotic term gets enhanced at large $E$, when $H$ becomes less important, implying that experiments working at larger $E$ are sensitive to smaller $\delta V_{\rm eff}$. We thus expect atmospheric and astrophysical data to be more constraining than reactor and solar experiments.

This may be confirmed comparing the analysis of atmospheric neutrinos presented in~\cite{Abe:2014wla} with the sensitivity of the DUNE experiment, that should give a measure of the potential of long baseline experiments. The Super Kamiokande collaboration reports bounds of order $|\delta V_{\rm eff}|<(0.5-4)\times10^{-23}$ GeV on $\mu e,\mu\tau, e\tau$ transitions.~\footnote{With the same dataset, constraints of comparable order of magnitude may actually be obtained on the diagonal terms $a_{\alpha\alpha}$ and $b$ as well, although these were not considered in that reference.} The previous bound corresponds to an effect of order less than $\sim(3-30)\%$ of the earth potential. The future sensitivity of DUNE on $a$ and NSI has been estimated in~\cite{Coloma:2015kiu,deGouvea:2015ndi}, and can be translated into $|\delta V_{\rm eff}|\lesssim10\%$ of the earth potential. This is in the ballpark of the current constraint from Super Kamiokande. Hence, existing atmospheric neutrino experiments already have a sensitivity comparable to the next generation of long baseline neutrino experiments, and we expect that future atmospheric data will only improve.~\footnote{For solar neutrinos the baseline is so large that the flux reaching the earth has completely lost its coherence and the new physics signature is encoded in an effective angle of order $\lesssim|\delta V_{\rm eff}|E/\Delta m^2_\odot\sim10^{-2}|\delta V_{\rm eff}|/(10^{-23}$ GeV$)$, resulting on a deviation from the SM that is too small to be detected in current and future facilities. We thus conclude that solar data are not so very promising either. Note that the authors of~\cite{Kostelecky:2003cr} reach a different conclusion because implicitly assume that the new physics effect is controlled by $\delta V_{\rm eff}L$. However, this is only true at short baselines.}

The smaller SM background at high $E$ is the reason why astrophysical neutrinos are often suggested as the best probes for Lorentz Violating interactions, see e.g.~\cite{Kostelecky:2013rv}. However, above a certain critical energy, neutrino oscillation experiments have structural limitations that cannot be removed by reducing the systematic or experimental uncertainties. Indeed, in the regime $\delta V_{\rm eff}\gg H$ the SM Hamiltonian is negligible, and the new physics actually dominates the transition:
\ba\label{largeE}
P(\alpha_\lambda\to\beta_{\lambda'})\propto\left(\frac{\delta V_{\rm eff, off}}{\delta V_{\rm eff}}\right)^2~~~~~~~~~~~~~~~({\rm for}~~~\delta V_{\rm eff}\gg H).
\ea
This might seem a welcome property because one can efficiently isolate the effects of the new physics. However, (\ref{largeE}) shows that the oscillation probability is now set by a ``dark angle," implying that the overall scale of $\delta V_{\rm eff}$ cannot be constrained. In particular, the magnitude of the effect crucially depends on the relative size of the flavor-violating and flavor-conserving entries in $\delta V_{\rm eff}$, and can be significantly small if the diagonal elements of $\delta V_{\rm eff}$ are sufficiently large. This represents an insurmountable limit for very energetic neutrinos. In fact, even in the presence of an ideal experimental precision and optimistically assuming the initial flavor content was known, one cannot set a robust constraint on the magnitude of $\delta V_{\rm eff}$ employing only data with $H\ll\delta V_{\rm eff}$. One can clearly see this limitation in a recent IceCube analysis of $E>400$ GeV~\cite{Aartsen:2017ibm}, where atmospheric neutrino data alone cannot set an upper bound on $\delta V_{\rm eff}$ unless further assumptions are made.

\begin{figure}[t]
\begin{center}
\includegraphics[width=10cm]{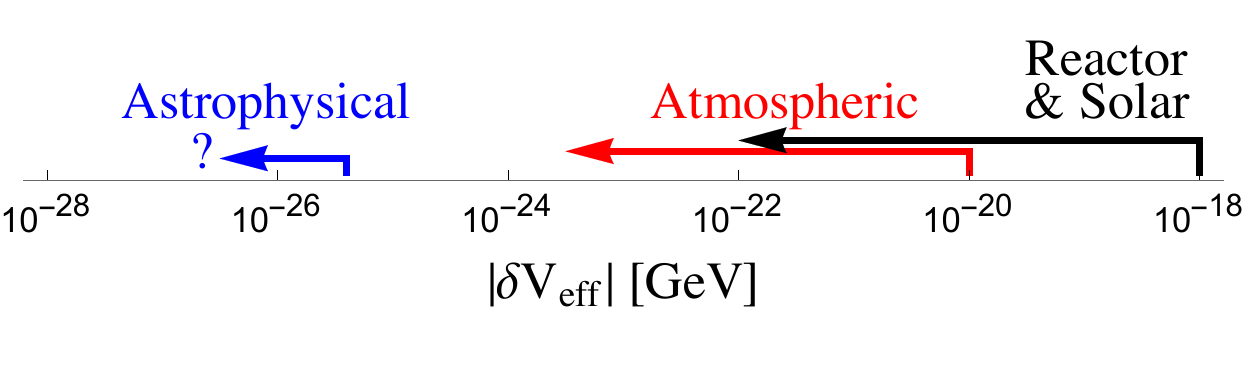}
\caption{Schematic plot illustrating the reach of astrophysical (blue) and atmospheric (red) neutrinos on the new matter potential, $|\delta V_{\rm eff}|$ (in GeV). The lower end of each region (indicated by an arrow) is set by the sensitivity of the experiment and can be lowered decreasing the systematic uncertainties. (We included a question mark in the ``Astrophysical" band because systematic uncertainties on the initial flavor content do not allow us to set limits at the moment.) On the other hand, the upper end --- indicated by a vertical segment --- is set by the requirement $|\delta V_{\rm eff}|<|H|$, as explained in the text, and is a structural constrain that cannot be overcome. Above that limit the experiment cannot probe the scale $\delta V_{\rm eff}$, only ratios of its elements. 
}\label{graph}
\end{center}
\end{figure}

The main message here is that it is only for energies satisfying $\delta V_{\rm eff}\lesssim H$ that we can, at least in principle, directly access the scale $\delta V_{\rm eff}$. The current situation is pictorially illustrated in Figure~\ref{graph}. Atmospheric data with $E<100$ GeV can be used to set a robust limit of order $|\delta V_{\rm eff}|\lesssim \Delta m_{\rm atm}^2/E\sim10^{-23}$ GeV. Reactors, accelerator and solar data have lower energies and are typically sensitive to larger exotic effects, see for instance Double-CHOOZ~\cite{Abe:2012gw} and MINOS~\cite{Adamson:2012hp}. Finally, astrophysical neutrinos have $E>30$ TeV and thus become fully effective only for $|\delta V_{\rm eff}|\lesssim4\times10^{-26}$ GeV. Hence, in order to reliably take advantage of the very high energy domain we first need to robustly close the gap $10^{-25}$ GeV $\lesssim|\delta V_{\rm eff}|\lesssim10^{-23}$ GeV by improving our atmospheric analysis.

%%%%%%%%%%%%%%%%%%
%%%%%%%%%%%%%%%%%%
%%%%%%%%%%%%%%%%%%
%%%%%%%%%%%%%%%%%%

Motivated by these considerations, we next analyze in detail the potential reach of two atmospheric neutrino experiments: ORCA and the high-energy IceCube. The former is intended to provide a measure of the discovery potential of the low energy spectrum ($E<100$ GeV), whereas the latter of the high energy tail ($E>100$ GeV) of atmospheric data.

\subsection{Analysis assumptions}

In general, $F$ and the vector component of $V$ define a preferred direction in space, and force both $a,b$ to depend on the orientation of the neutrinos. A dedicated comparison among experiments would thus need to take into account the different orientations of the neutrino beams of the various facilities as well as seasonal variations. Since our paper is focussed on the prospects of current and future experiments, it would be impossible to include such effects properly. For this reason we will assume 
\ba\label{Ass}\label{15par}
a(\hat{\bf p})=\left(
\begin{matrix}
a_{ee} & a_{e\mu} & a_{e\tau} \\ 
a_{e\mu}^* & a_{\mu\mu} & a_{\mu\tau} \\
a_{e\tau}^* & a_{\mu\tau}^* & a_{\tau\tau}
\end{matrix}
\right),~~~~~~~~~~~~~~
b(\hat{\bf p})=\left(
\begin{matrix}
0 & b_{e\mu} & b_{e\tau} \\ 
-b_{e\mu} & 0 & b_{\mu\tau} \\
-b_{e\tau} & -b_{\mu\tau} & 0
\end{matrix}
\right)
\ea
are independent of ${\bf\hat p}$ throughout our analysis. This simplifying assumption is violated if a sizeable portions of the neutrino flux has significantly different directions, or if the duration of the experiment is long enough to appreciate seasonal effects. Still, because these latter conditions are quite special, we believe our assumption (\ref{Ass}) is capable of capturing the most generic signatures of these scenarios. We will verify this claim a posteriori below.

Furthermore, a priori we have no reason to prefer a certain coupling over another. It is however not practically possible, nor particularly useful, to scan over all 15 parameters (\ref{Ass}). We therefore decide to present our projections under the assumption that {\emph{one complex coupling}} at a time is turned on. Besides simplicity, we believe this approach is defendable for the following reasons. First, phenomenologically viable scenarios have exotic potentials $\delta V_{\rm eff}$ of a magnitude smaller than the SM effective potential. As a consequence, transition probabilities proportional to more powers of $a,b$ are suppressed compared to processes with a single $a,b$ insertion. %For example, in scenarios with $b\neq0$ (see Appendix~\ref{sec:examples}) one finds that $P(\nu_\alpha\to\bar\nu_\beta)$ is larger than  $P(\nu_\alpha\to\nu_{\beta\neq\alpha})$ by a factor $|b|/V_{\rm SM}$.~\footnote{We note in passing that atmospheric data currently do not discriminate between neutrinos and anti-neutrinos, so that neutrino/anti-neutrino transitions are not qualitatively different from ordinary oscillations and a fair comparison between $P(\nu_\alpha\to\bar\nu_\beta)$ and $P(\nu_\alpha\to\nu_{\beta\neq\alpha})$ makes sense.} 
Second, while correlations among the various entries in $a$ and/or $b$ are expected to be suppressed, there can still be non-trivial cancellations between the real and imaginary parts of a given $a_{\alpha\beta}$ which we find useful to show explicitly. This is emphasized in the $a_{\mu\tau}$ example in Appendix~\ref{sec:examples}. We we do not anticipate that this feature applies to other $a$s or to $b$s, but for consistency (and as a check of our expectations) we will keep both real and imaginary parts of all couplings.

The reader should keep in mind that turning on more than one coupling can reduce sensitivity to a given coupling. This degeneracy has been previously seen in oscillation studies of NSI (e.g.~\cite{Friedland:2005vy,Liao:2016orc}) and persists for the $a$ and $b$ couplings introduced here.

\subsection{Details on the numerical analysis}
We report here the approach used for simulating both ORCA and IceCUBE experiments.

\subsubsection{ORCA}
The analysis of the prospective constraints of ORCA on $a$ and $b$ is performed analogously to what done in \cite{Capozzi:2017syc} for the mass ordering. Here we only modify the calculation of the oscillation probability in order to accommodate the propagation of neutrinos in a generic background. For the sake of brevity we report below only the most important features of the analysis. Further details are found in \cite{Capozzi:2017syc}. 

ORCA \cite{Adrian-Martinez:2016fdl} is a new low-energy neutrino telescope (part of the KM3NeT project) based
on the Cherenkov detection technique in seawater, offshore from Toulon, France. The experiment is dedicated to studying atmospheric neutrinos of all flavors in the energy range [2,100] GeV. We focus here on upward-going neutrinos, i.e. with zenith angle $\theta\in$ [$\pi/2,\pi$] where $\theta=\pi/2$ ($\pi$) represents the horizontal (vertical) direction.

The calculation of the event spectra in ORCA consists of three parts: production, propagation and detection.
Concerning neutrino production we use the fluxes calculated in \cite{Honda:2015fha} for the Frejus site, assuming no mountain overburden. Regarding neutrino propagation, we divide the Earth into five shells and in each we calculate the evolution operator up to the second order in the Magnus expansion \cite{Fogli:2012ua}. Finally, all the ingredients relevant for simulating neutrino detection are taken from \cite{Adrian-Martinez:2016fdl}, considering the configuration with 9 meters of vertical spacing between the optical modules. In particular, each neutrino event can be reconstructed either as a track or as a cascade, depending on its topology. We therefore divide the events according to these two categories.

 We define $N_{i,j}^\alpha(p_k)$ as the number of simulated experimental data events of type $\alpha$ (track, cascade) in the $ij$-th bin, where $i$ and $j$ represent the indices for energy and angular bins, respectively. $p_k$ are the current best fit values of the oscillation and systematic parameters, with $k$ an index labelling them. Note that we generate the mock experimental data $N_{i,j}^\alpha(p_k)$ assuming no non-standard effect in the propagation of neutrinos, i.e. $a=b=0$. We define $\tilde{N}_{i,j}^\alpha(a,b,\tilde{p}_k)$ as the number of expected number events for a set of oscillation and systematic parameters ($\tilde{p}_k$), in general different from their best fit, and either $a\ne0$ or $b\ne 0$. We calculate the $\Delta\chi^2$ marginalizing over systematic and oscillation parameters $\tilde{p}_k$. This minimization is performed through the pull method \cite{Fogli:2002pt}
 %%%%
\begin{equation}
\Delta\chi^2(a,b) = \min_{\tilde p_k}\left[  \sum_{i=1} \sum_{j=1} \sum_{\alpha=\text{track,cascade}}\frac{\left(N^\alpha_{ij}(p_k)-\tilde N^\alpha_{ij}(a,b,\tilde p_k)\right)^2}{N^\alpha_{ij}(p_k)+(\sigma_uN_{ij}^\alpha(p_k))^2} + \sum_k\left(\frac{p_k-\tilde p_k}{\sigma_k}\right)^2\right],
\label{chi2}
\end{equation}
where $\sigma_k$ is the current uncertainty on oscillation and systematic parameters and  $\sigma_u$ represents an uncorrelated uncertainty in each bin.
We adopt the default set of systematic uncertainties proposed in \cite{Capozzi:2017syc}, i.e. including $\sigma_u=0.015$ and a 1.5\% residual uncertainty on the shape of the events distributions, which we parametrize as a quartic polynomial as in \cite{Capozzi:2017syc}. All other sources of systematics with their correspondent uncertainties can be found in \cite{Capozzi:2017syc}.

\subsubsection{IceCUBE}

IceCUBE is a neutrino telescope in the underground ice of the South Pole based on the Cherenkov detection technique. Despite being designed mainly for ultra-high energy neutrinos, i.e. those with energy greater than $\sim20$ TeV, it performs also measurements of the atmospheric neutrino flux in the range  [0.3,20] TeV.

The analysis method adopted for IceCube is similar to the one described for ORCA and summarized in Eq. \ref{chi2}. However, in this case we do not simulate the prospective mock data $N_{i,j}^\alpha(p_k)$. Instead, in order to obtain any given exposure, we rescale the observed experimental data contained in a recent data release \cite{TheIceCube:2016oqi}, 20145 muons collected in 343.7 days. The calculation of the expected number of events $\tilde{N}_{i,j}^\alpha(a,b,\tilde{p}_k)$ is performed by folding Monte-Carlo weigths, provided in \cite{TheIceCube:2016oqi}, with the atmospheric neutrino flux and oscillation probability. The same data release contains information for the implementation of some systematic uncertainties.
In particular, we consider a 2.5\% error on the $\nu/\bar{\nu}$ ratio in the atmospheric neutrino flux, 10\% on the ratio $\pi/K$, 40\% on the total normalization and 5\% on the cosmic ray spectral index. Regarding detection we are marginalizing over  the optical modules efficiency and different ice models, where the first is a continuous variable and the second is discrete. Finally we use the Honda+Gaisser neutrino flux given in the data release associated with Ref.~\cite{TheIceCube:2016oqi}.

The data contained in \cite{TheIceCube:2016oqi} seems to contain a statistical fluctuation which enhances the constraints on $a$ and $b$ (about a factor of 2) with respect to expectations for the same exposure. To obtain stronger constraints we therefore decided to rescale the observed distribution of events to obtain the mock data for a generic exposure, rather than simulating the data as done for our ORCA analysis.

%%%%%%%%%%%%%%%%%%%
%%%%%%%%%%%%%%%%%%%
\begin{figure*}[h!]
\begin{center}
 \includegraphics[width=.49\textwidth]{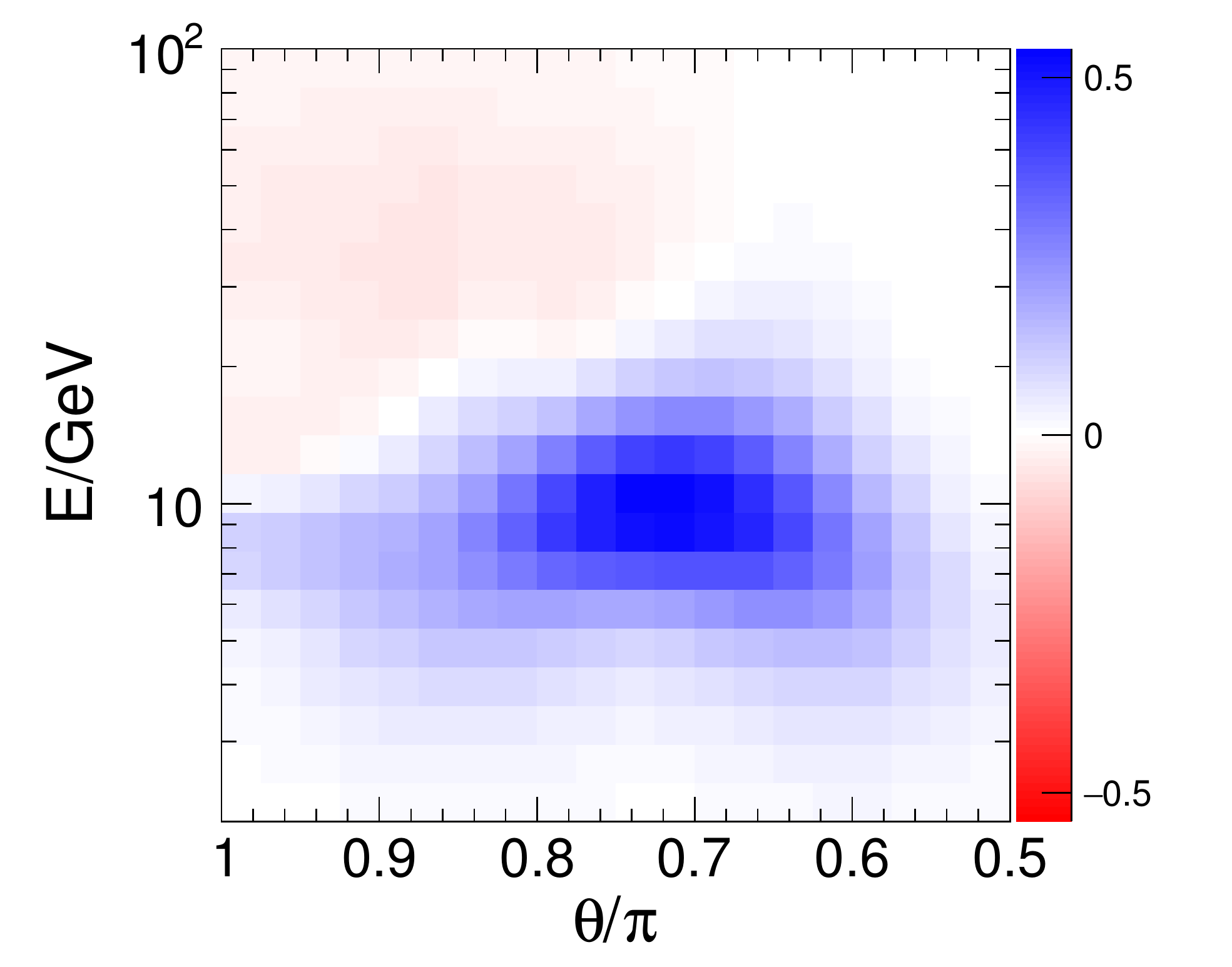}
        \includegraphics[width=.49\textwidth]{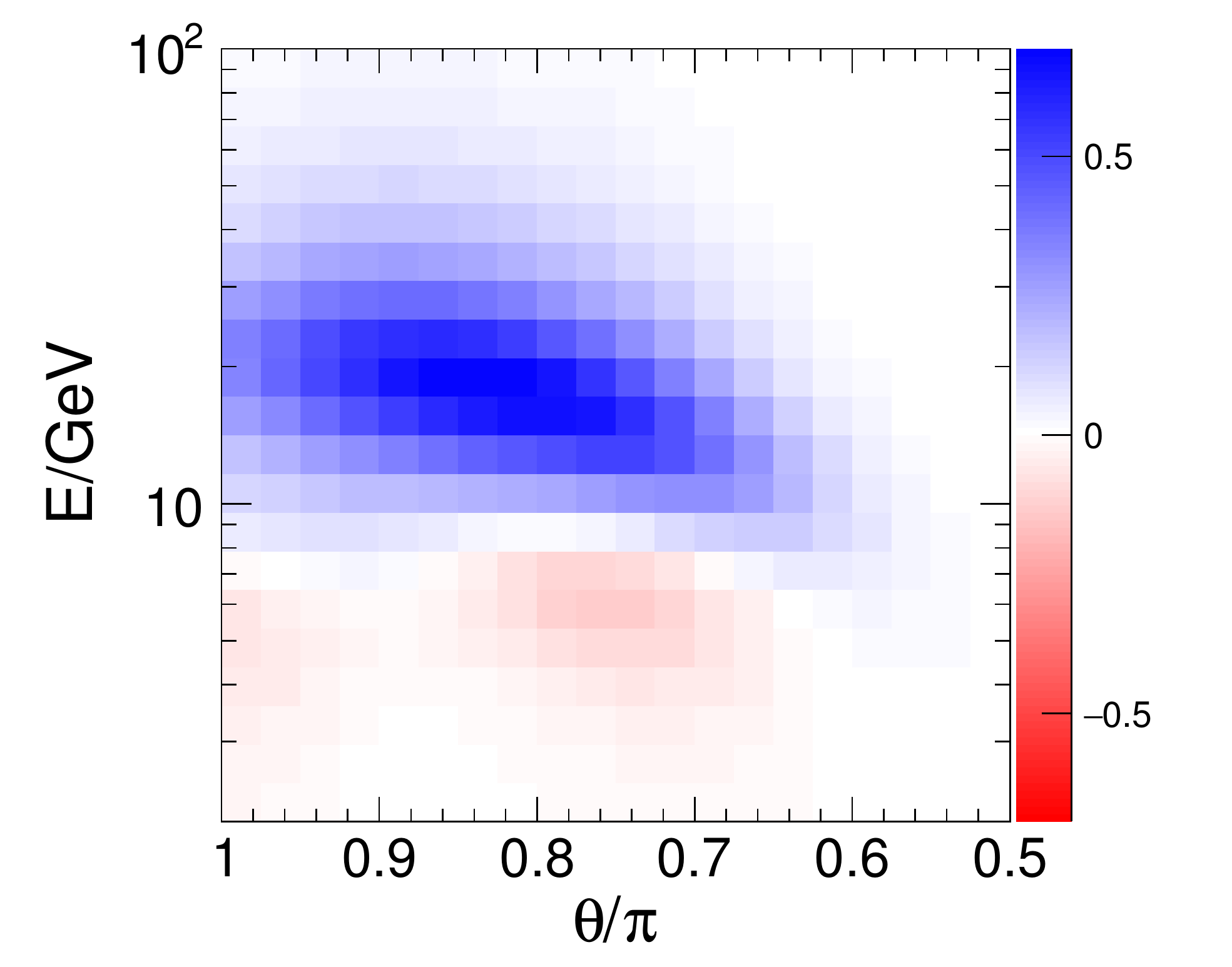}    \\
 	\includegraphics[width=.49\textwidth]{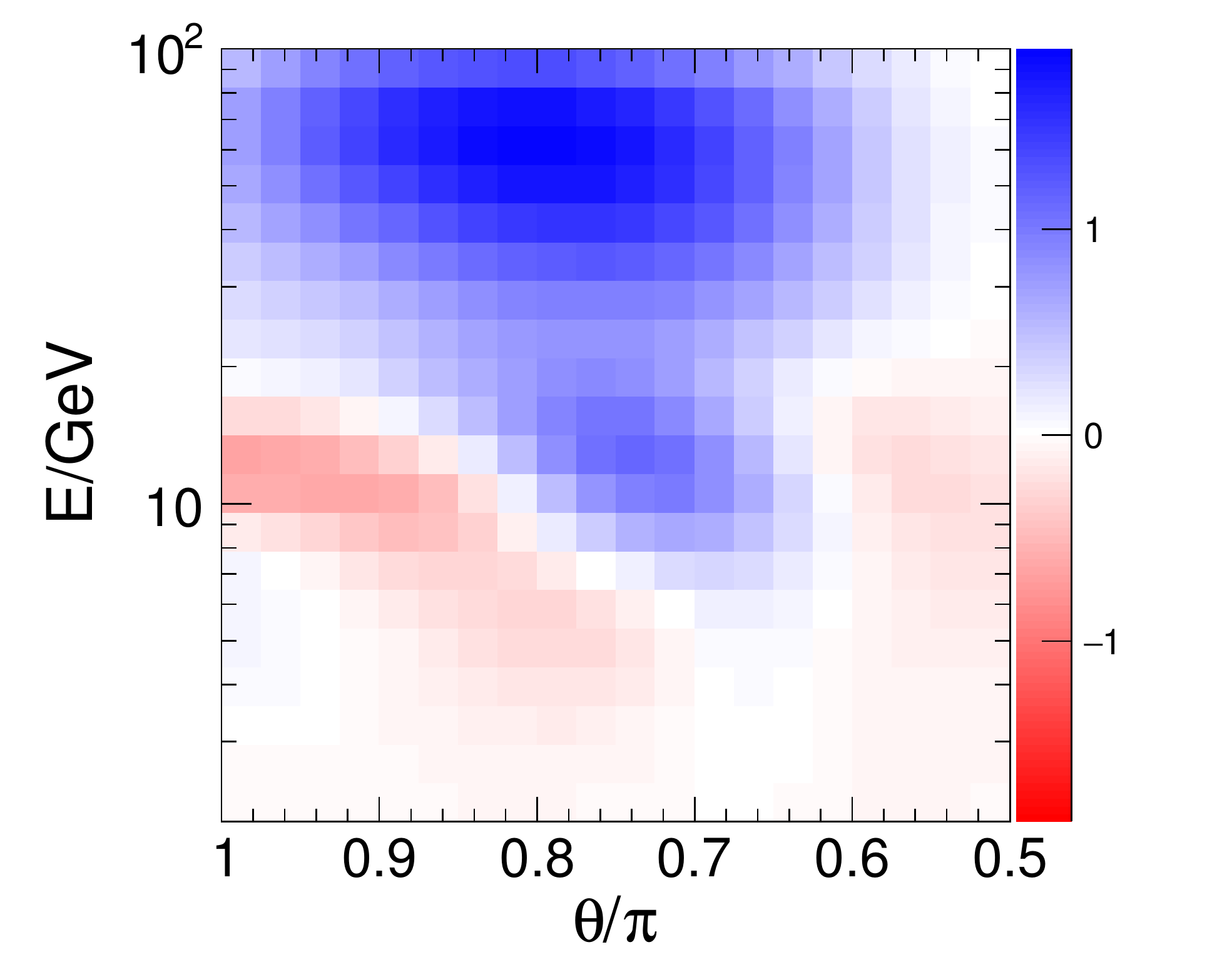}
        \includegraphics[width=.49\textwidth]{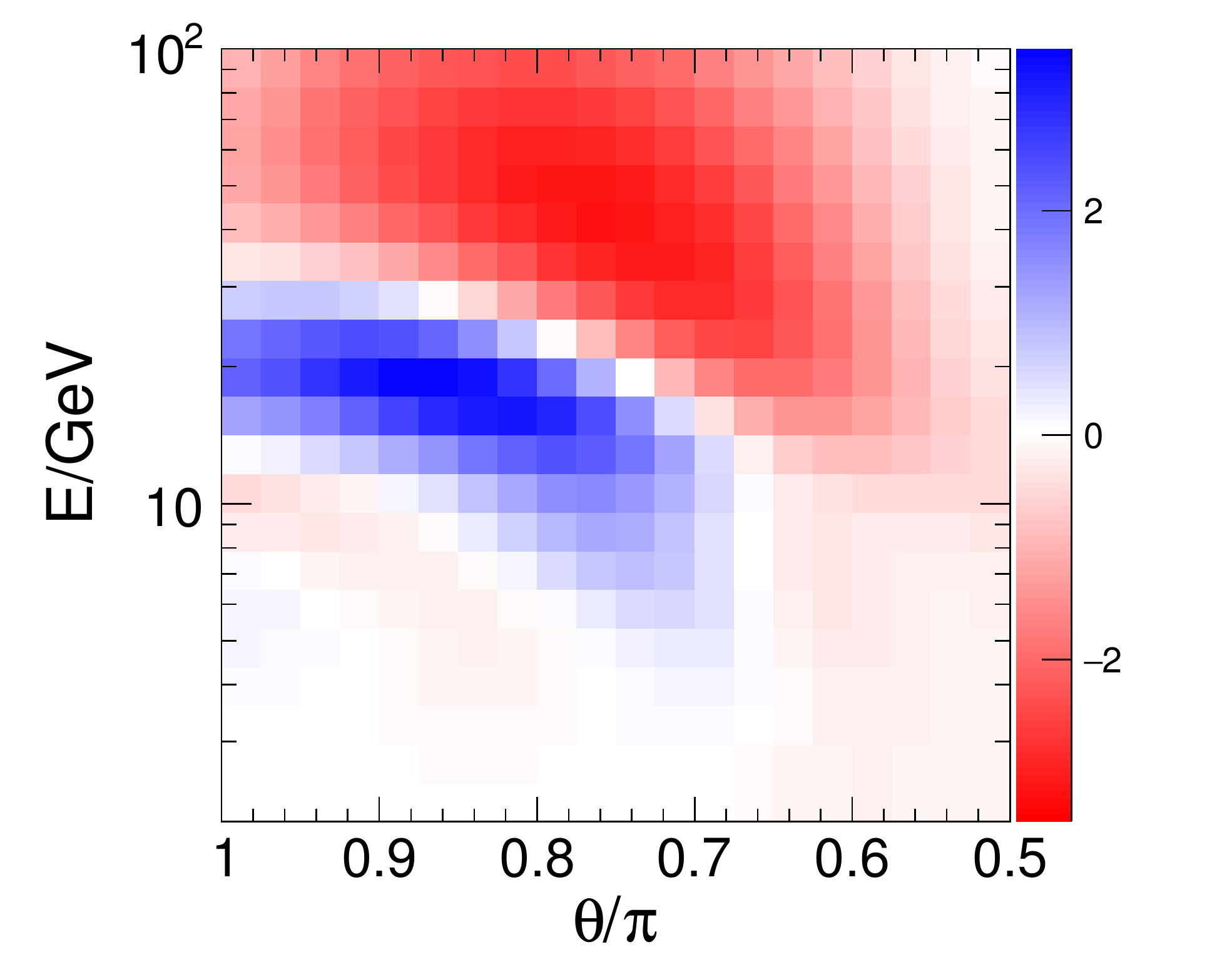}    \\
        \includegraphics[width=.49\textwidth]{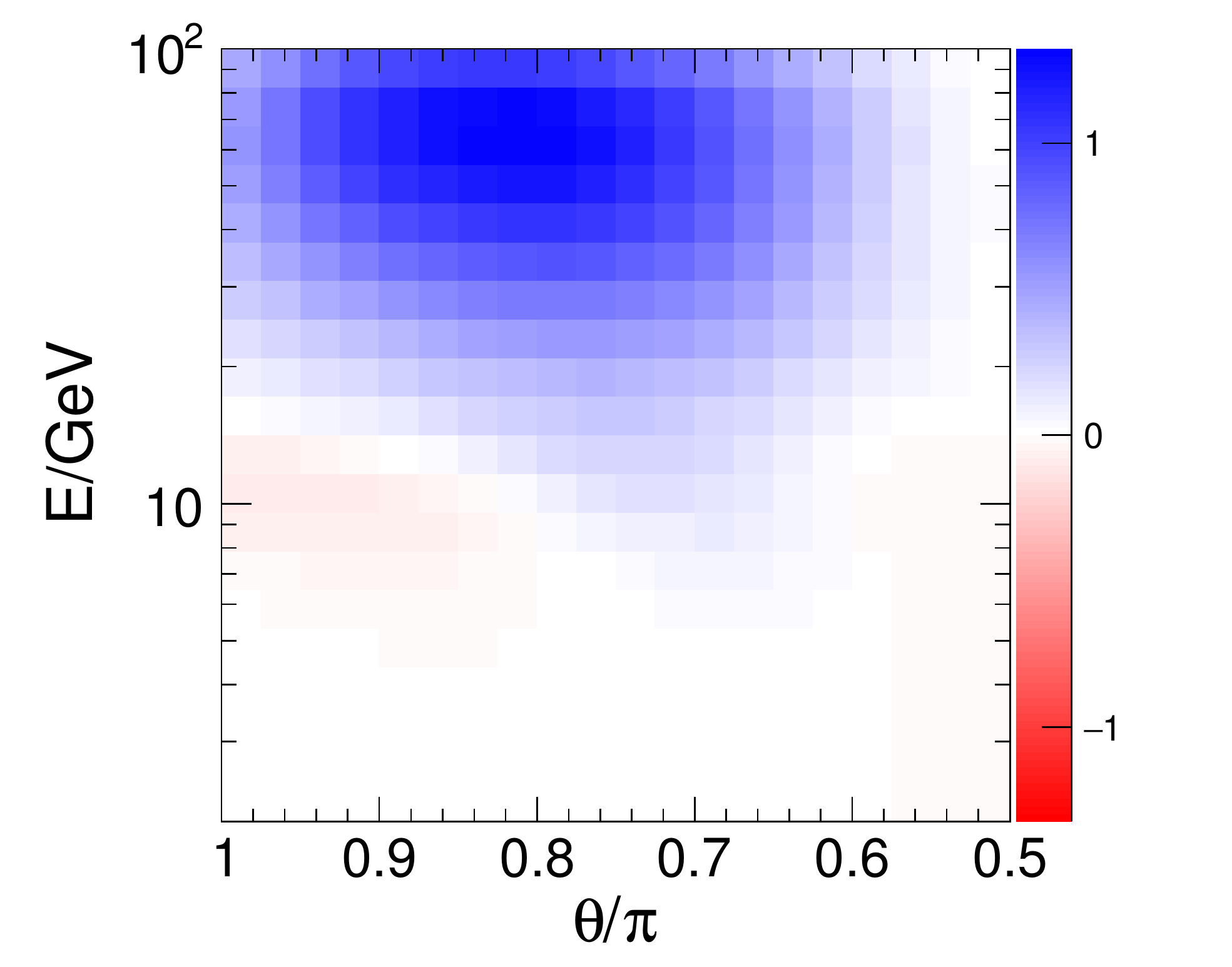}
        \includegraphics[width=.49\textwidth]{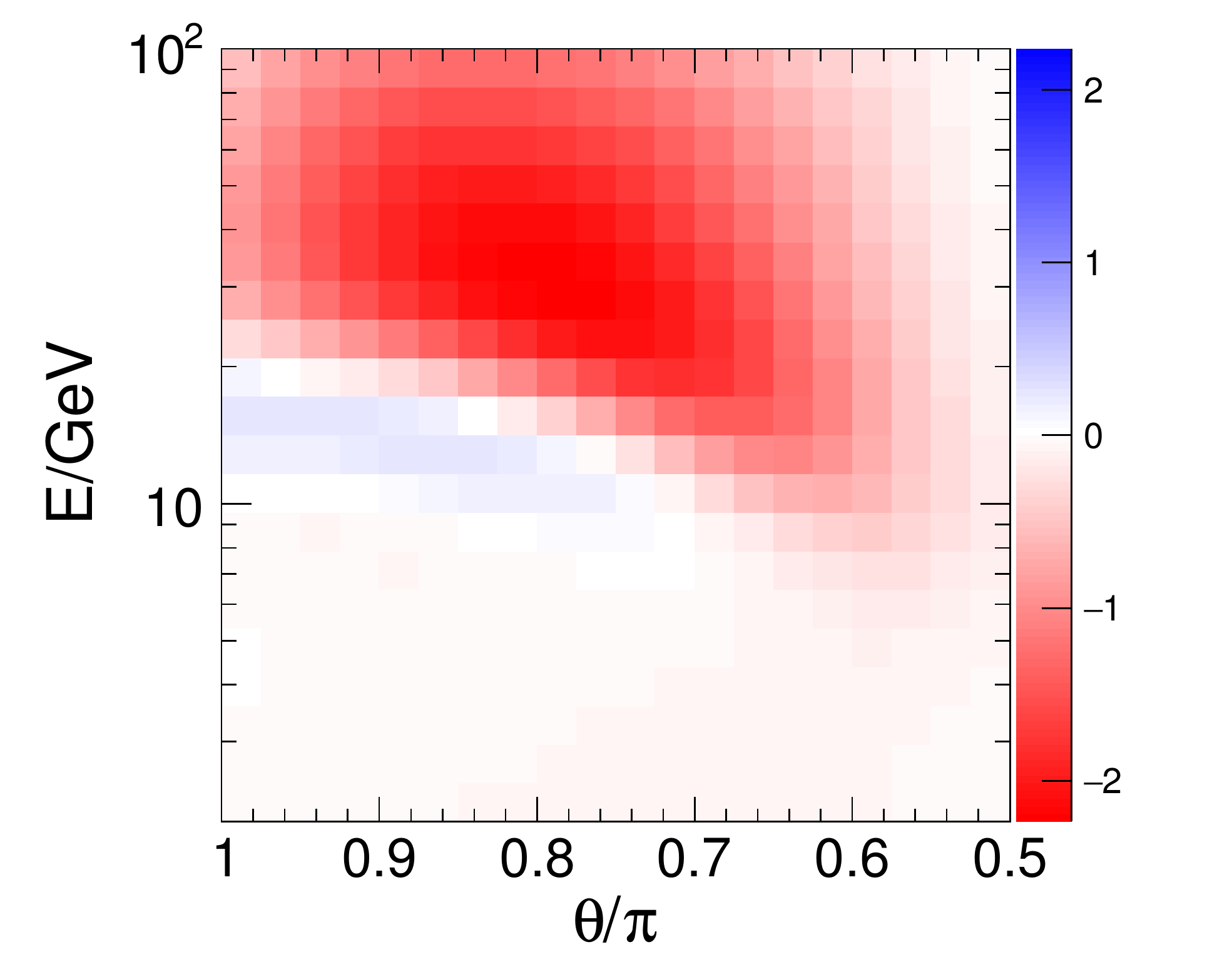}    \\
\caption{Here we display $(N-N_{\rm SM})/\sqrt{N_{\rm SM}}$ distributions for 10 years of data at ORCA in cascades ({left}) and tracks ({right}) as they depend on zenith angle and energy, where $\theta=\pi/2$ ($\pi$) represents the horizontal (vertical) direction. {Top panel}: The only nonzero coupling is $a_{\mu \mu} = 10^{-23}~{\rm GeV}$, and {middle panel} we have $a_{\mu \tau} = 10^{-23}$ GeV. {Bottom panel}: Here we take $b_{\mu \tau} = 10^{-23}~{\rm GeV}$. In these examples, the normal mass hierarchy is chosen for definiteness. %\com{Note that these broad distributions imply that accounting for the directionality of the dark background would result in similar sensitivities.} 
}
\label{fig:oscillo}
\end{center}
\end{figure*}
%%%%%%%%%%%%%%%%%%%
%%%%%%%%%%%%%%%%%%%

\subsection{Results}
\label{sec:numerical}

To get a physical sense of the types of modifications dark media can have on neutrino oscillations we display in Fig.~\ref{fig:oscillo} the deviations from the SM expectation as it varies with energy and zenith angle $\theta$. The estimator of such a deviation is taken to be $(N-N_{\rm SM})/\sqrt{N_{\rm SM}}$, the difference between number of events with and without $a,b$ divided by the square root of the SM prediction. As a representative sample we consider $a_{\mu \mu} = 10^{-23}~{\rm GeV}$ (top panel), $a_{\mu \tau} = 10^{-23}$ GeV (middle panel), and  $b_{\mu \tau} = 10^{-23}~{\rm GeV}$ (bottom panel).

Before discussing each case in turn, it is useful to make some introductory qualitative remarks. First, at high $E$ the flux is mostly composed of $\nu_\mu,\bar\nu_\mu$; in track processes we mostly see $\mu$ neutrinos, whereas cascade events include all flavors. Second, high $E$ effects are dominated by new physics and maximized at longer paths. For this reason we expect the main SM deviations to peak towards $\theta\sim\pi$. At low $E\sim10$ GeV the SM is sizable and the deviations are closer to $\theta\sim\pi/2$, where the SM oscillation length peaks, $L_{{\rm osc}} \sim 10^{3}~{\rm km}~(E/{\rm GeV})$.

These qualitative observations allow us to interpret the panels in Fig.~\ref{fig:oscillo}. First consider the top panel of Fig.~\ref{fig:oscillo}, showing the effect from $a_{\mu \mu} \neq 0$. As in all cases with diagonal $a$, this parameter {\emph{requires}} interference with the SM in order to produce physical effects, resulting in a peak at relatively moderate energies. This is seen in both the track and cascade rates. By contrast, off-diagonal $a,b$ do not rely on the SM and their effect is usually maximized at large $E$, when the SM contribution is suppressed. This is shown in the central and lower panels of Fig.~\ref{fig:oscillo} for $a_{\mu \tau} \neq0$ and $b_{\mu \tau} \neq0$, respectively. Still, $a_{\mu \tau} \neq0$ can also benefit from interference with the SM, unlike the $b_{\mu \tau} \neq0$ case, thus explaining why the intermediate plots contain both low- and high-energy features in the distribution. The color coding is also consistent with physical expectations: $(a,b)_{\mu\tau}$ tend to remove tracks (i.e. $\nu_\mu$) and generate cascades (i.e. $\nu_\tau$ in this case). Similar considerations apply to $a_{e\mu}, b_{e\mu}$. Finally, we observe that all of the distributions exhibit rather broad deviations in zenith angle, $\theta$. This justifies a posteriori our assumption that a more detailed analysis including directional dependence in the matter potential would not lead to very different distributions or sensitivities.

%Finally we now examine a case with nonzero $b$, turn on $b_{\mu \tau} \neq0$. In this case now there is no interference and we see that compared to the $a_{\mu \tau}$ example the $\sim 10$ GeV peaks are removed. As a result we expect weaker sensitivity to $b_{\mu \tau}$ than to $a_{\mu \tau}$. 

%%%%%%%%%%%%%%%%%%
%%%%%%%%%%%%%%%%%%
\begin{figure}[h!]
\begin{center}
\includegraphics[scale=0.78]{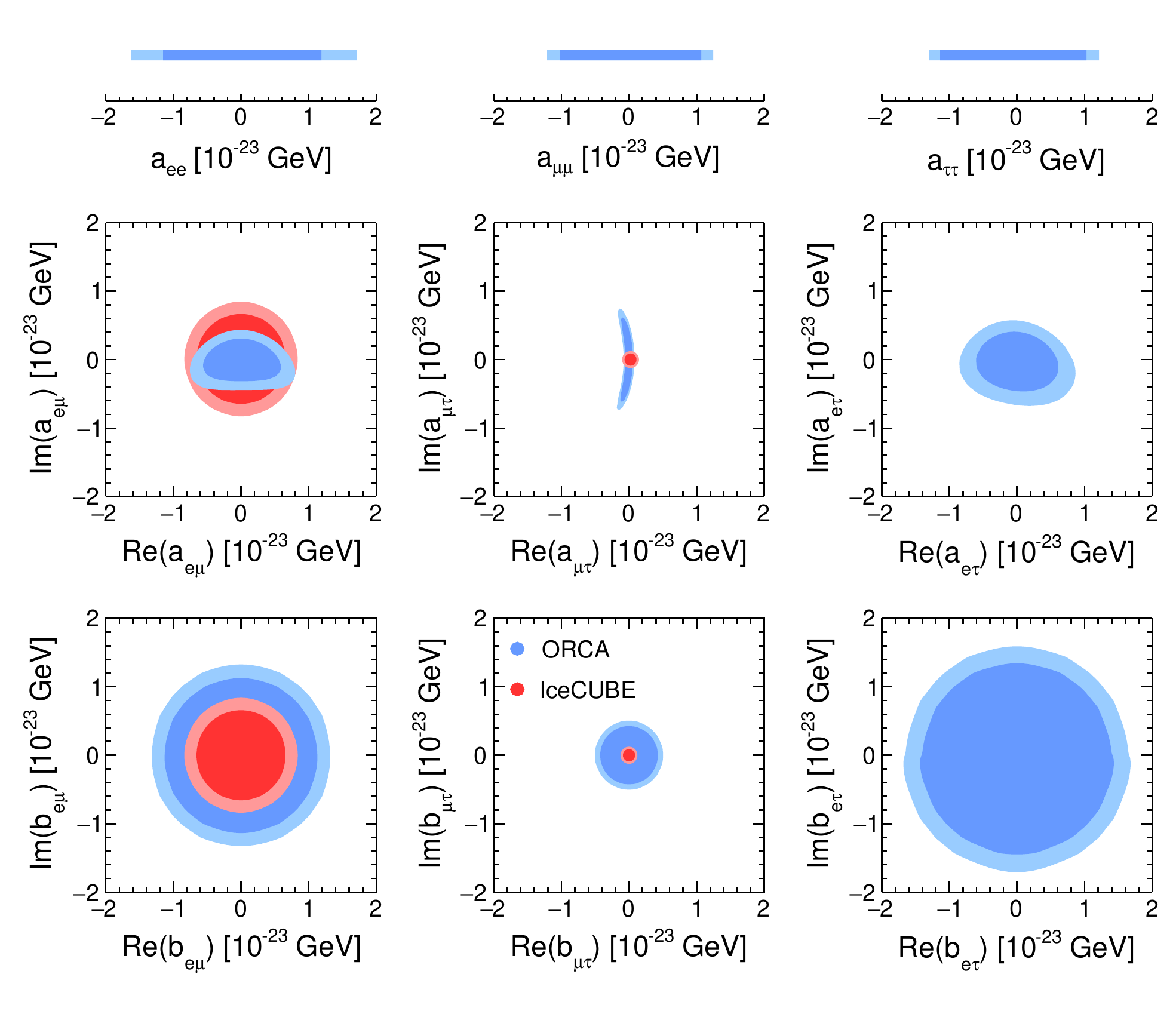}%~~\includegraphics[width=8cm]{figs/parameter_scan_dm2_sol_s2_12_SNO_Be7_pep_neg.pdf}
\caption{Here we summarize our projected 10 year sensitivities at IceCube (high-energy and muon only) and ORCA to dark backgrounds with nonzero $a$ and $b$ couplings each turned on one at a time. {Note that $b$ couplings have no interference effects with the SM so we expected circular regions. The same is true for all IceCube sensitivities which are done at energies so large that the SM contribution is absent. These bounds assume the NH (see Fig~\ref{fig:limitsIH} for bounds in the IH). }
}\label{fig:limitsNH}
\end{center}
\end{figure}
%%%%%%%%%%%%%%%%%%
%%%%%%%%%%%%%%%%%%

%%%%%%%%%%%%%%%%%%
%%%%%%%%%%%%%%%%%%
\begin{figure}[h!]
\begin{center}
\includegraphics[scale=0.78]{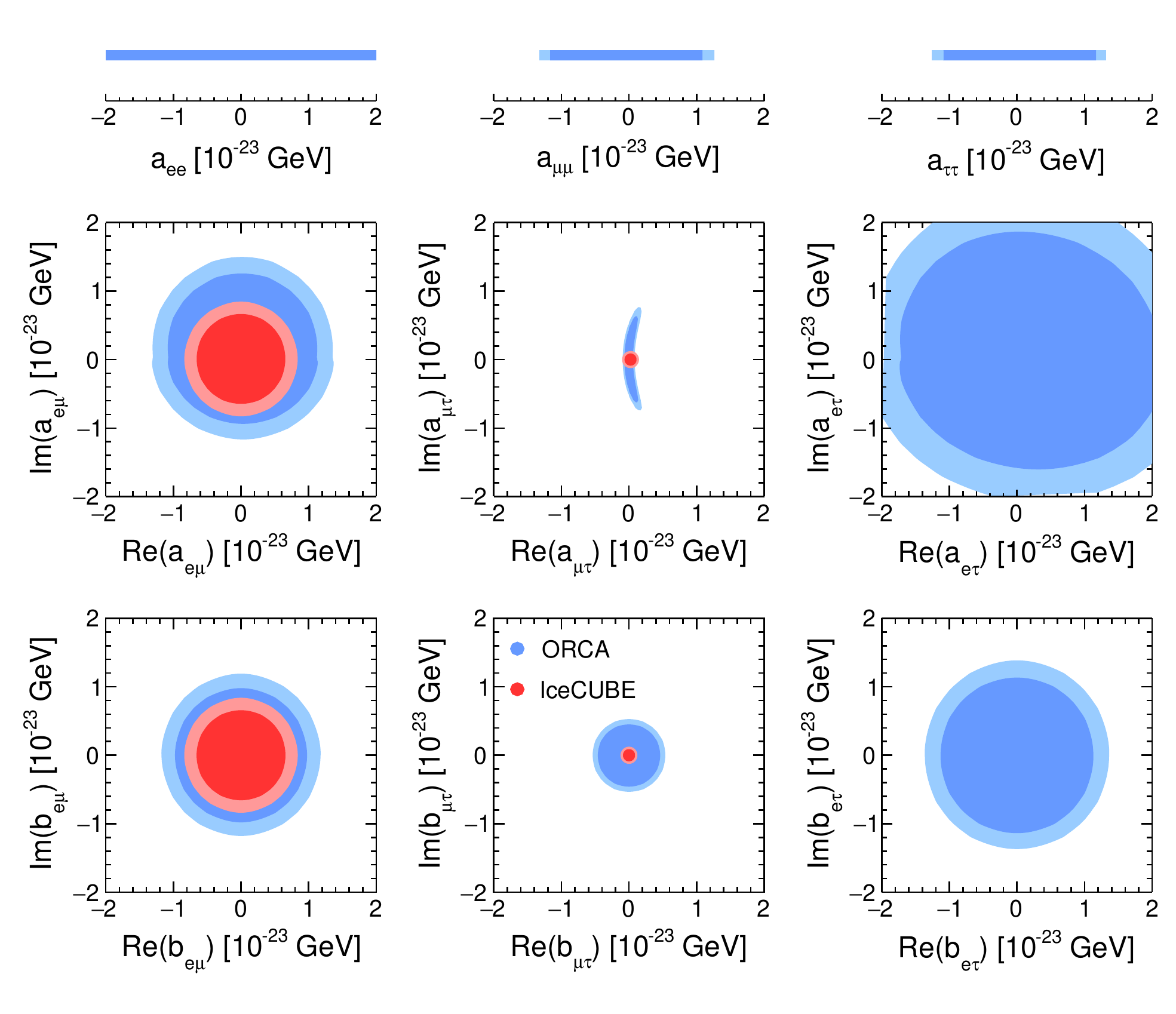}%~~\includegraphics[width=8cm]{figs/parameter_scan_dm2_sol_s2_12_SNO_Be7_pep_neg.pdf}
\caption{Same as Fig.~\ref{fig:limitsNH} but assuming IH.
}\label{fig:limitsIH}
\end{center}
\end{figure}
%%%%%%%%%%%%%%%%%%
%%%%%%%%%%%%%%%%%%

Let us now turn to the future sensitivity estimates we make for the IceCube and ORCA experiments displayed in Fig.~\ref{fig:limitsNH} (for normal hierarchy, NH) and Fig.~\ref{fig:limitsIH} (for inverted hierarchy, IH). On the upper panels we show the sensitivity to flavor diagonal $a$ couplings. As discussed above, since these couplings require interference with the SM to be observable our high-energy IceCube analysis is not relevant. Here the low-energy sensitivity of ORCA is therefore critical. Moreover, the sensitivity is roughly similar for all three neutrino flavors, {with $a_{\mu \mu, \tau\tau}$ being somewhat more sensitive owing to the larger $\nu_{\mu}$ flux, large $\theta_{23}$ angle, and better flavor sensitivity in the track channel.}

{Regarding the flavor off-diagonal couplings of Figs.~\ref{fig:limitsNH} and \ref{fig:limitsIH} we observe that $a$ is usually more constrained than $b$ whenever $a$ appears linearly in the oscillation probability (see (\ref{scalingP}) and (\ref{scalingP'})). On the other hand $b$ always appears quadratically, see also Appendix~\ref{sec:examples}.} This latter point also explains why all allowed regions in $b$ are defined by circles of constant $|b|$. A related feature that the figures reveal is that the IceCube constraints are symmetric in ${\rm Im}[a], {\rm Re}[a]$, whereas those from ORCA are not in general. This again follows from the fact that the SM contribution at the energies relevant to ORCA is not negligible. Consider the $a_{\mu \tau}$ coupling for definiteness, for which an approximate expression for the oscillation probability is derived in Appendix~\ref{sec:examples}, see eq. (\ref{Pcanc}). There it is seen explicitly that only the real part of $a_{\mu \tau}$ appears linearly in (\ref{scalingP}), and that for moderate energies the contribution of the real and imaginary parts may partially cancel leading to non-trivial sensitivity in the complex plane. This explains the ``banana" shaped contours in Figs.~\ref{fig:limitsNH} and \ref{fig:limitsIH}. 

{The main difference between Normal Hierarchy and Inverted Hierarchy is that in the latter case the SM neutrino conversions of neutrinos are less efficient, while the antineutrino transitions are enhanced but suffer from smaller statistics. Hence, the constraints on the parameters that take advantage of the SM (as emphasized above, this happens only at ORCA energies) are relaxed, see $a_{ee,e\mu, e\tau}$ in Figs.~\ref{fig:limitsNH} and~\ref{fig:limitsIH}. Also, the ``banana" shape is reversed because $\Delta m^2_{\rm atm}$ changes sign.} 

In general we observe a very rich structure with both low and high energies contributing in a non-trivial way to event distributions (see the middle panel of Fig.~\ref{fig:oscillo}). This richness leads to IceCube being stronger in some cases ($a_{\mu \tau}, b_{\mu\tau,e\mu}$) and ORCA being stronger in the others ($a_{e\tau,e \mu}, b_{e\tau}$). It is also worth recalling that our IceCube analysis relies only on muon neutrinos, so that the $e-\tau$ transitions at IceCube are not probed.

%Finally for the $b$ couplings (necessarily off-diagonal), there can be no SM interference since $b$ contributes to $\nu_{\alpha} \rightarrow \bar{\nu}_{\beta \neq \alpha}$.  Therefore all sensitivities should be the same in the real and imaginary directions. Moreover, since there is no SM contribution the highest energies should be most constraining (as seen too in the bottom panel of Fig.~\ref{fig:oscillo}), and we expect IceCube to set the strongest constraints. 

%\begin{figure}[t]
%\begin{center}
%\includegraphics[scale=0.6]{figs/constraints_diagonal_a.pdf}%~~\includegraphics[width=8cm]{figs/parameter_scan_dm2_sol_s2_12_SNO_Be7_pep_neg.pdf}
%\caption{sensitivity... 
%}\label{diagonal_constraints}
%\end{center}
%\end{figure}
%%%%%%%%%%%%%%%%%%%
%%%%%%%%%%%%%%%%%%%

%{\bf Important caveat to this ``one-at-a-time'' analysis. There is an ambiguity/degeneracy since the limits are weaker when turn on several couplings. This happens in NSI too.}

\section{Models}
\label{sec:models}

Now we turn our attention to realistic models with neutrino couplings to dark sector states. We will examine various implications of additional neutrino-DM interactions and provide a consistency check on the framework. We will see that in the present scenarios there are large regions of parameter space where neutrino oscillation modifications set far stronger limits than any other physical process.

Phenomenological considerations similar to the ones presented in this section have been previously presented in~\cite{Capozzi:2017auw} in an Asymmetric DM model. There it was emphasized that a very efficient way to generate effective couplings for active neutrinos --- without spoiling the physics of the charged leptons --- is via the mixing with exotic fermions $\nu_s$ that couple to the dark sector~\cite{Capozzi:2017auw}. For definiteness we will assume this mechanism is at work in all the models we discuss. We will not consider cosmological constraints on the sterile sector, however. These can be very constraining, but are in general model-dependent. To be on the safe side we will assume cosmology is standard because the active-sterile mixing started at very late stages of the Universe's history~\cite{Patwardhan:2014iha,Vecchi:2016lty}. 

\subsection{Scalar $m$} 
\label{sec:scalar}
%%%
A correction to $m$ can only lead to observable effects if it possesses some space-time dependence. Otherwise, $m$ would be indistinguishable from a neutrino mass. 

To realize a space-time varying $m$ we need a scalar dark field $\phi$ with a space-time dependent background. Consider scenarios in which $\phi$ is a classical field oscillating around the origin of field space with energy density $\rho_\phi\sim m_\phi^2\phi^2$. If some coupling to neutrinos (active or sterile) exists, say $y\phi \nu_s\nu_s$, one can obtain an effective space-time-varying mass of order
\ba
m\sim \sin^2\theta_s~y\frac{\sqrt{\rho_\phi}}{m_\phi},
\ea
with $\sin\theta_s$ the active/sterile mixing. Note that, consistently with the general scaling argument presented in the introduction, this has the structure $\sim\sqrt{G_{\rm D}\rho_{\rm D}}$ with $G_{\rm D}\sim \sin^2\theta_sy^2/m_\phi^2$. This possibility has been considered recently in~\cite{Berlin:2016woy,Krnjaic:2017zlz,Brdar:2017kbt}. The authors find that such effect may result in observable variations in neutrino oscillations if $\rho_\phi$ is of the order of the local dark matter density, $y/m_\phi\gtrsim1$ eV, and the oscillation frequency $1/m_\phi$ is not much larger nor smaller than years.

In concrete examples, however, the non-derivative couplings of $\phi$ must be very small. Otherwise, sizable $y$ couplings would generate large quantum corrections to the scalar potential, both at zero and finite temperature. The former should be minimized in order to avoid a large fine-tuning. This is often achieved assuming $\phi$ is an approximate Nambu-Goldstone boson, in which case $y\lesssim m_\phi/f_\phi$ with $f_\phi$ some large scale. With this expectation, however, neutrino experiments would be unable to detect any effect in the foreseen future. Yet, if we ignore the fine-tuning problem and allow $y\gg m_\phi/f_\phi$, we still have to deal with large thermal effects that can dramatically impact the evolution of the scalar. For example, thermal corrections of order $m_\phi^2(T)\sim y^2T^2$ would prevent the scalar from behaving as cold dark matter until $yT<m_\phi(0)$. Because we experimentally know that dark matter was already present at $T\sim1$ eV, one can derive --- barring miraculous coincidences --- a structural constraint $y/m_\phi(0)\ll1/$ eV that is stronger than the projected sensitivity of future neutrino experiments. $\phi$ may also be identified with a subdominant dark matter component or even radiation, but in these cases its density would be too small to have a sizable effect.

Our conclusion is that $m$ is basically indistinguishable from a constant mass term unless a large amount of fine-tuning is invoked. We will therefore focus on $V,F$ in the rest of the paper.

%%%%%%%%%%%%%%%
\subsection{Vector $V$} 
%%%%%%%%%%%%%%%

%{\color{blue}Interactions with ordinary matter at rest are parametrized at leading order by a non-vanishing $V^0$~\cite{Wolfenstein:1977ue}\cite{Mikheev:1986gs} (the effect of ${\bf V}\neq0$ was discussed for example in~\cite{Nunokawa:1997dp} and~\cite{Diaz:2009qk}).} 

In the following we consider the possibility of a dark $V^\mu$. We focus on scenarios in which the latter is generated by a fundamental vector field, but the reader should keep in mind that $V^\mu\propto\partial^\mu\phi$ is another interesting possibility.

\subsubsection{Model 1: Asymmetric Dark Matter Coupled to an Ultra-Light Vector}

In a previous paper~\cite{Capozzi:2017auw} we discussed a simple realization of the vector scenario in which an asymmetric dark matter background with density $J^\mu$ sources a spontaneously broken dark vector field $A_\mu$ coupled to (sterile) neutrinos, thus generating 
\ba\label{V1}
V^\mu&=&\sin^2\theta_s\frac{g^2_A}{m^2_A} J^\mu\\\no
&\sim&10^{-24}~{\rm GeV}~\sin^2\theta_s\left(\frac{g_A~{\rm keV}}{m_A}\right)^2\left( \frac{J^\mu m_{\rm DM}}{0.4~{\rm GeV}/{\rm cm}^3}\right)\left(\frac{\rm keV}{m_{\rm DM}}\right), 
\ea
with $\sin\theta_s$ the active/sterile mixing ($\sin^2\theta_s=1$ if a direct coupling between active neutrinos and the dark world exists), and $g_A,m_A$ the vector coupling and mass. In the second line of (\ref{V1}) we estimated the magnitude of the current assuming it is due to a dark matter candidate of mass $m_{\rm DM}$ and matter density $\rho_{\rm DM}\sim J^\mu m_{\rm DM}\sim0.4$ GeV$/$cm$^3$. Hence we find the scaling $\delta V_{\rm eff}\sim G_{\rm D}\rho_{\rm D}/m_{\rm D}$. Importantly, while weakly-coupled sectors $g_A\ll1$ cannot be directly probed otherwise, they can still show up in neutrino oscillation experiments if $G_{\rm D}\sim \sin^2\theta_sg_A^2/m_A^2$ is large enough.

 {A quantitative discussion of the astrophysical and cosmological impact of this picture can be found in~\cite{Capozzi:2017auw}. In that work it was found that IceCube and SN1987A can place stringent bounds at low mediator masses by inducing a neutrino mean free path for scattering on the cosmic neutrino background. At larger mediator masses, $m_A\gtrsim 10^{-2}~{\rm eV}$, bounds on DM kinetic decoupling in the early universe from elastic scattering on the relativistic neutrinos become the most stringent non-oscillation bounds. All these become ineffective as $g_A$ decreases, however.}

\subsubsection{Model 2: Ultra-Light Vector DM} 
\label{sec:ul}

An additional model for vectorial DM interactions with neutrinos was introduced in~\cite{Brdar:2017kbt}, as an extension of the model of vector DM  sourced by inflation in~\cite{Graham:2015rva}. For definiteness, we assume that the only relevant couplings of the vector is to sterile neutrinos as in~\cite{Capozzi:2017auw}
\be \mathscr{L} \supset g_{A}~ \bar{\nu}_{s} \gamma_{\mu} \nu_{s} A^{\mu},
\ee
where $g_{A}$ is the new gauge coupling of the theory. We leave both the mass scale of the steriles $m_{s}$ and their mixing angles with the active neutrinos unspecified. Note that unlike~\cite{Brdar:2017kbt} we do not explore gauged lepton number models but instead focus on models where the sterile sector is charged under the new gauge symmetry. %
%Exotic backgrounds may arise if active or sterile neutrinos couple to a new gauge force with a non-trivial charge density.

 %(Couplings due to scalars, say $\partial_\mu\phi$ will be ignored because to avoid fine-tuning the latter are all suppressed by a large scale $f_\phi$, as mentioned above, and thus are too suppressed...)

%(In~\cite{Capozzi:2017auw} we made the additional assumption that the dark matter can be trapped in the sun due to other interactions with our world. We will not make this assumption here. Hence, the present paper deals with more generic signatures of the model.)

In this case it is the VEV of the vector, $\langle A \rangle \sim \sqrt{\rho_{\rm DM}}/m_{A}$, that induces a matter potential
\ba\label{V2}
V^\mu&=&\sin^2\theta_s\,g_A A^\mu \\ \no
&\sim&10^{-24}~{\rm GeV}~\sin^2\theta_s\left(\frac{g_A}{5\times10^{-19}}\right)\left(\frac{10^{-6}~{\rm eV}}{m_A}\right)\sqrt{\frac{\rho_{\rm DM}}{0.4~{\rm GeV}/{\rm cm}^3}}.
\ea
The potential now scales as $\delta V_{\rm eff}\sim\sqrt{G_{\rm D}\rho_{\rm D}}$, with $G_{\rm D}\sim \sin^4\theta_sg_A^2/m_A^2$. 

Note that sizable effects in neutrino oscillations are obtained with an extremely small coupling $g_A$. Because of this, all other constrains can be evaded, as emphasized in the introduction. Additional constraints here mainly come from: (1) dangerous early Universe processes which can equilibrate neutrinos and DM, %{\color{red} is this different?} 
(2) early Universe freeze-in~\cite{Hall:2009bx} of DM, and (3) present-day bounds on the neutrino mean free path from SN1987A~\cite{Kolb:1987qy} and IceCube~\cite{Cherry:2014xra,Cherry:2016jol}. We find all of these to be many orders of magnitude weaker than the oscillation bounds. The details of these constraints are outlined in Appendix~\ref{sec:constraints}.

\subsection{Tensor $F$} 
%%%%%%%%%%%%%%%

Tensor couplings parametrized by $F_{\mu\nu}$ are induced by dipole interactions with a dark gauge boson $A^\mu$ whose electric or magnetic fields are non-trivial. This interaction is expected to dominate over a possible $V^\mu$ background if the gauge symmetry associated to $A^\mu$ is unbroken. Indeed, gauge invariance may forbid a direct interaction between (active or sterile) neutrinos and $A^\mu$, in which case $F^{\mu\nu}$ would parametrize the leading effect.

Neutrino oscillations mediated by an electro-magnetic dipole moment were first studied in~\cite{Cisneros:1970nq}. Here we will allow the potential to be entirely due to dark forces. In practice, our tensor coupling parametrizes the interactions between neutrinos and a ``Dark CMB".

To estimate the size of $F$, we imagine adding exotic neutrinos $\nu_s$ that mix with the active ones as in~\cite{Capozzi:2017auw} and have a Yukawa coupling $g_s\nu_s\Psi\Phi$ with two particles $\Psi,\Phi$ charged under the dark radiation. This framework predicts
\ba
F_{\mu\nu}&\sim&\sin^2\theta_s\frac{g_s^2}{16\pi^2}\frac{m_s}{M^2}g_AA_{\mu\nu}\\\no
&\sim&10^{-24}~{\rm GeV}~{\sin^2\theta_s}\left(\frac{g_s}{10^{-1}}\right)^2\left(\frac{g_A}{10^{-1}}\right) \left(\frac{100~{\rm eV}~m_s}{M^2}\right) \sqrt{\frac{{\rho_{\rm D,rad}}}{1~{\rm eV/cm}^3}}.
\ea
where we take the current Dark CMB density to be $\rho_{\rm D, rad} \sim A_{\mu\nu}^2$, whereas $M$ is the largest mass of the two mediators $\Psi,\Phi$, and finally $g_A$ is the gauge coupling. Note that despite the rather small values of $M,m_s$, the effective operator description remains valid in our oscillation setup up since the oscillation impact comes from scattering at negligible momentum transfer. In this case the dark background is again scaling as $\delta V_{\rm eff}\sim\sqrt{G_{\rm D}\rho_{\rm D}}$, though now $G_{\rm D}\sim\sin^4\theta_s({g_s^2}/{16\pi^2})^2{g^2_Am_s^2}/{M^4}$. 

Similarly to the vector models, we find that the most constraining non-oscillation data comes from SN1987A. Assuming negligible relic abundances for $\Psi,\Phi$ (which may have annihilated away into dark radiation), the neutrino flux is depleted via $\nu_a + \bar\nu_{a,s} \rightarrow ({\rm anything})$, where $\nu_{a,s}$ are active or sterile relic neutrinos. %Moreover, the working fiducial assumption is that the $\Psi, \Phi$ particles are so light that they can be produced directly in the collision, $\nu + \bar\nu \rightarrow \Psi + \bar\Psi$ and  $\nu + \bar\nu \rightarrow \Phi + \Phi^*$, with total cross sections scaling like $\sigma \propto g_s^4/s$. In this case we find that the bound on the coupling is quite weak, $g_{s} \lesssim 0.3$. 
We estimate that the total cross section for the latter processes scales at most like $\sigma \sim g_s^4/s$. Requiring the mean-free paths for the neutrinos emerging from the supernova is long enough we find a rather weak bound on the coupling, $g_{s} \lesssim 0.3$. Again, oscillation experiments provide the dominant constraints in the weak-coupling, low-mass regime.

\section{Conclusions}

Neutrinos acquire exotic interactions in many extensions of the Standard Model. The effect of non-standard couplings to ordinary matter has been investigated extensively in the literature, both in the context of oscillation experiments as well as colliders. These are usually very severely constrained by rare processes involving charged leptons and quarks. Here we considered the possibility that the new couplings involve a dark sector. Such a picture is far less constrained and neutrino experiments can play a pivotal role. In fact, the dark energy, dark matter, and possibly dark radiation that permeates the world around us might act as a medium affecting neutrino propagation in a non-trivial way while still being undetected otherwise: neutrino oscillation experiments represent a unique opportunity to detect light dark sectors interacting with our world via the neutrino portal.

Given our ignorance on the nature of dark sector physics we have considered the most general classes of backgrounds: scalar, vector (parametrized by a hermitian matrix $a$), and tensor (parametrized by an antisymmetric matrix $b$). Neutrino oscillations in the dark medium can be described via an effective ``Hamiltonian" that we derived under the hypothesis that the exotic fields are nearly constant within the relevant space-time scales, and induce subdominant effects on top of ordinary neutrino oscillations. The true origin of the leading neutrino oscillations is not crucial to our numerical analysis, and may be due to ordinary masses or even the dark backgrounds themselves. While the statement that a dark {\emph{scalar}} background can be responsible for neutrino oscillations comes hardly as a surprise, we showed that even {\emph{vector and tensor}} backgrounds with anisotropies at small scales may in fact {\emph{mimic}} vacuum oscillations without the need to invoke neutrino masses nor chiral symmetry breaking.

Neutrino forward scattering is affected by an effective potential $\delta V_{\rm eff}$ that scales with appropriate powers of $G_{\rm D}\rho_{\rm D}$, where $G_{\rm D}\sim g_{\rm D}^2/m_{\rm D}^2$ is a dark ``Fermi scale" of the exotic sector and $\rho_{\rm D}$ is the dark background energy density. Neutrino oscillation experiments are unique probes of scenarios with small $g_{\rm D},m_{\rm D}$ and sizable $G_{\rm D}$, when all other probes become inefficient. 

Models in which the dark background is generated by a density current have $\delta V_{\rm eff}\sim G_{\rm D}\rho_{\rm D}/m_{\rm D}$, as in the SM, whereas when the dark density is generated by the vacuum expectation value of bosonic fields one finds $\delta V_{\rm eff}\sim\sqrt{G_{\rm D}\rho_{\rm D}}$. In either case a determination of $\delta V_{\rm eff}$ is necessary to extract useful information on the scale of the dark sector. We have argued that $\delta V_{\rm eff}$ can only be constrained by experiments that have the ability to probe the energy range in which $|\delta V_{\rm eff}|\lesssim|H|$, with $H$ indicating the SM effective Hamiltonian. On the other hand, for too large energies the SM background becomes irrelevant ($H\ll\delta V_{\rm eff}$) and we lose access to the magnitude of $\delta V_{\rm eff}$. For this reason atmospheric neutrinos turn out to be a necessary input for other, more energetic probes. To close the gap schematically shown in figure~\ref{graph}, a careful analysis of atmospheric data is necessary. 

We analyzed the sensitivity of the ORCA detector and the IceCube experiment to dark backgrounds. We find that these experiments have impressive sensitivity, especially in the $\mu$-$\tau$ sector, and an interesting complementary, with IceCube's high-energy sensitivity setting the leading constraints on $a_{\mu \tau}$, $b_{\mu \tau}$, and $b_{e\mu}$ and ORCA providing stronger bounds on $a_{e \mu}$, $a_{ee}$, $a_{\mu \mu}$, $a_{\tau \tau}$, $a_{e\tau}$, and $b_{e\tau}$.

The atmospheric neutrino flux we have focused on has important limitations however. A major drawback is the poor $\nu/\bar\nu$ discrimination power, which inevitably impacts the sensitivity to the tensor interaction $b$. Moreover, atmospheric neutrinos are mostly sensitive to muons, with $e\tau$ transitions being only weakly constrained at present. Long-baseline experiments such as DUNE will therefore be crucial in the search of dark backgrounds $a,b$. Based on the estimates made in~\cite{Coloma:2015kiu,deGouvea:2015ndi}, for certain couplings DUNE may have sensitivity competitive with atmospheric data. The possibility of better $\nu/\bar{\nu}$ discrimination may aid DUNE in improving the constraints especially on tensor backgrounds not considered in~\cite{Coloma:2015kiu,deGouvea:2015ndi}. 

%{\color{red} comparison to NSI? }

\section*{Acknowledgments} IMS is very grateful to the University of South Dakota for support. F.C. acknowledges partial support by NSF Grant PHY-1404311 to J.F. Beacom, the Deutsche Forschungsgemeinschaft
through Grant No. EXC 153 (Excellence Cluster "Universe"), Grant No.
SFB 1258 (Collaborative Research Center "Neutrinos, Dark Matter,
Messengers") as well as by the European Union through Grant
No. H2020-MSCA-ITN-2015/674896 (Innovative Training Network "Elusives"). The work of LV is supported by the Swiss National Science Foundation under the Sinergia network CRSII2-160814.

\newpage

\appendix

\section{Derivation of (\ref{effH})}
\label{app:}

We here aim to derive $P$ in the presence of small backgrounds $\delta m=m-m_\nu, V,F$, with $m_\nu$ the unperturbed mass. To proceed we make three simplifying assumptions, that are motivated in the text:
\begin{itemize}
\item[{\bf(1)}] the  $m,V,F$ do not violate space-time translations;
\item[{\bf(2)}] $\delta m,V,F$ do no contain space-time derivatives acting on $N$;
\item[{\bf (3)}] $V,F$ can be interpreted as small perturbations. 
\end{itemize}
With the first assumption, the full energy-momentum operator $P^\mu=P^\mu_0+\delta P^\mu$ is conserved, and we are allowed to write $P^\mu[N(t,{\bf x})]=P^\mu[N(0,{\bf x})]$. Here and in the following we denote by $P^\mu_0$ the 4-momentum operator in the absence of $\delta m, V,F$ (but with a mass $m_\nu$). Then, defining $N_0(t,{\bf x})\equiv e^{-iH_0t}N(0,{\bf x})e^{+iH_0t}$ --- with the subscript indicating that this field satisfies the free field equations --- we have $P^\mu[N(t,{\bf x})]=e^{+iH_0t}P^\mu[N_0(t,{\bf x})]e^{-iH_0t}$.

Assumption (2) implies that $\delta P^i=0$ and, together with $[H_0,P_0]=0$, allows us to obtain the important result $P^i=P_0^i$. For simplicity, since $m, m_\nu$ become essentially indistinguishable, we simply set $m=m_\nu$ in $P^\mu_0$ and forget about $\delta m$ in the following. Note that a standard calculation tells us that in the ultra-relativistic regime $H_0=P_0+\delta_mH$, with $\delta_mH$ a small correction. From this latter definition and the first two assumptions immediately follows that, for a given neutrino direction:
\ba\label{Pgen}
P=P_0=H_0-\delta_{m} H=H-e^{+iH_0t}\delta H_{\rm int}(t)e^{-iH_0t}-\delta_{m} H,
\ea
where $\delta H_{\rm int}(t)=-\int d{\bf x}~\delta{\cal L}[N_0,N^\dagger_0]$ represents the term induced by $V,F$ expressed in the {\emph{interaction picture}}.

Finally, assumption (3) allows us to expand in the perturbations and obtain
\ba\label{Ppert}
P&=&H-e^{+iH_0t}[\delta_{m} H+\delta H_{\rm int}(t)]e^{-iH_0t}\\\no
&=&H-e^{+iHt}[\delta_{m}H+\delta H_{\rm int}(t)]e^{-iHt}+{\cal O}(\delta H^2)\\\no
&=&e^{+iHt}[H-\delta_{m}H-\delta H_{\rm int}(t)]e^{-iHt}+{\cal O}(\delta H^2).
\ea
While $P$ is time-independent, $\delta H_{\rm int}(t)$ in general does depend on $t$. However, the above form is very useful in practice because $\delta H_{\rm int}(t)$ is straightforward to compute under the first two assumptions: we just need to calculate $\delta{\cal L}$ on a solution of the free field equations associated to ${\cal L}_0$.

We are now in a position to obtain our result (\ref{effH}). To introduce the notation that we will adopt in the following, we write $H_0$ in terms of the creation operators $a^\dagger_{\pm,\alpha}$ of neutrinos of a given helicity ($-$ for particles and $+$ for anti-particles) and flavor index $\alpha=e,\mu,\tau,\rm etc$. Ignoring terms of ${\cal O}(m^4/|{\bf p}|^3)$, a standard calculation gives
\ba\label{dm}
H_0&=&\int\frac{d{\bf p}}{(2\pi)^3}(a_-^\dagger~~a_+^\dagger)
\left.\left(
\begin{matrix}
|{\bf p}|+\frac{mm^*}{2|{\bf p}|} &  \\ 
 & |{\bf p}|+\frac{m^*m}{2|{\bf p}|}
\end{matrix}
\right)\right|_{p_0=|{\bf p}|}
\left(
\begin{matrix}
a_- \\ 
a_+
\end{matrix}
\right)+{\cal O}(m^4)\\\no
&\equiv&P_0+\delta_{m} H.
\ea
The flavor indices have been suppressed for simplicity. The reader should interpret $a_\pm,m$ as matrices in flavor space. With our conventions, the operators $a_{\pm,\alpha}({\bf p})$ carry helicity and flavor indices and generate 1-particle states $\sqrt{2E}a^\dagger|0\rangle$. The creation/annihilation operators satisfy $\left\{a_\lambda,a_{\lambda'}^\dagger\right\}=\delta_{\lambda,\lambda'}(2\pi)^3\delta^{(3)}({\bf p}-{\bf p}')$.

Because in (\ref{Ppert}) we are ignoring mixed terms involving products of $m,V,F$, the free field $N_0$ used to determine $\delta H_{\rm int}(t)$ can be taken to be a solution of the massless equation:
\ba\label{free}
N_0(x)=
\left.\int\frac{d{\bf p}}{(2\pi)^3}\left[\xi_-e^{-ip\cdot x}a_{-}+i\sigma_2\xi_+^*e^{+ip\cdot x}a_{+}^\dagger\right]\right|_{p_0=|{\bf p}|}+{\cal O}(m)
\ea
with $\xi_-$ and $\xi_+=-i\sigma_2\xi_-^*$ a complete set of spinors. They are functions of ${\bf p}$ with negative and positive helicity, $p^i\sigma^i\xi_\pm=\pm\frac{1}{2}|{\bf p}|\xi_\pm$. Inserting (\ref{free}) in $\delta{\cal L}$ we finally obtain:
\ba\label{Vint}
\delta H_{\rm int}(t)&\equiv&-\int d{\bf x}~\delta{\cal L}[N_0,N^\dagger_0]\\\no
&=&\int\frac{d{\bf p}}{(2\pi)^3}(a_-^\dagger~~a_+^\dagger)
\left.\left(
\begin{matrix}
-\frac{p^\mu V_\mu}{|{\bf p}|} & \frac{4\sqrt{2}}{|{\bf p}|}{p}^\mu(F_{\mu\nu})\epsilon^\nu \\ 
\frac{4\sqrt{2}}{|{\bf p}|}{p}^\mu(F_{\mu\nu})^\dagger(\epsilon^\nu)^* & +\frac{p^\mu V_\mu^*}{|{\bf p}|}
\end{matrix}
\right)\right|_{p_0=|{\bf p}|}
\left(
\begin{matrix}
a_- \\ 
a_+
\end{matrix}
\right)\\\no
&+&{\cal O}(aa,a^\dagger a^\dagger).
\ea
Again, flavor indices have been suppressed for brevity. In obtaining (\ref{Vint}) we used $\xi_-^\dagger\bar\sigma^\mu\xi_-=\xi_+^\dagger\sigma^\mu\xi_+={p^\mu}/{|{\bf p}|}$, the completeness relation $\xi_-\xi_-^\dagger+\xi_+\xi^\dagger_+={\bf 1}_{2\times2}$, and defined $\xi_-^\dagger\sigma^\mu\xi_+=-\xi_-^\dagger\bar\sigma^\mu\xi_+\equiv\sqrt{2}\,\epsilon^\mu$. From the previous definition follows that $\epsilon^\mu=(0,\vec\epsilon)$ satisfies $\epsilon^0=0$, $\epsilon^\mu p_\mu=0$, $\epsilon^\mu\epsilon^*_\mu=-1$. The latter polarization tensor was previously used in~\cite{Kostelecky:2003cr}. 

%For relativistic neutrinos the full Hamiltonian can be written as $H=P+\delta H$, where $\delta H$ takes into account the small corrections to the formula $E=|{\bf p}|$ induced by $m,V,F$. In field theory one calculates $P$ by finding a solution $N$ of the equations of motion and plugging it into.... 

Eq.~(\ref{Vint}) is the final result; from it one can obtain (\ref{effH}) observing that $|\nu(x)\rangle\equiv e^{-i\delta H x}|\nu(0)\rangle$, with $\delta H=\delta_{m}H+\delta H_{\rm int}(t)$ is defined in terms of (\ref{dm}) and (\ref{Vint}). A few comments are in order. First, as mentioned at the beginning of this section, oscillation probabilities depend on the amplitudes $\langle\nu'|e^{iPL}|\nu\rangle$, with initial and final states $|\nu\rangle, |\nu'\rangle$ constructed via linear combinations of $a_{\pm,\alpha}^\dagger$ with the same energy (but in general different 3-momentum) and well-defined flavor and helicity quantum numbers. The factors of $e^{\pm iHt}$ in (\ref{Ppert}) thus become immaterial phases and neutrino propagation is effectively controlled by $\delta H+{\cal O}(\delta H^2)$.

Second, the operator $\delta H_{\rm int}(t)$ depends explicitly on $t$. However, this dependence disappears from the oscillation probabilities, consistently with (\ref{Pgen}). Indeed, the time dependence is relegated to the ${\cal O}(aa,a^\dagger a^\dagger)$ terms of~(\ref{Vint}). These have vanishing matrix elements with $|\nu\rangle,|\nu'\rangle$ (that are linear combinations of $a^\dagger_{\pm,\alpha}|0\rangle$) and can be neglected in the study of neutrino oscillations.

Finally, note that the same procedure employed to derive (\ref{Vint}) may be used to obtain a matter potential in the case of slightly varying backgrounds. The result would be the same as in (\ref{Vint}) as long as the length scale within which the potentials vary is much longer than $1/|{\bf p}|$. In that case one can split the calculation in various regions of approximately constant backgrounds, where (\ref{Vint}) holds. In particular, our result with $V^\mu=V^0({\bf x}), F^{\mu\nu}=0$ reproduces the standard matter potential of~\cite{Wolfenstein:1977ue}\cite{Mikheev:1986gs}.

\section{Constraints on the ultra-light vector dark matter model}
\label{sec:constraints}

Here we briefly summarize some of the bounds on the ultra-light vector dark matter model of Sec.\ref{sec:ul}, beyond neutrino oscillations. We will see that these are substantially weaker for sub-eV mediators. 

Having set the scale of masses and couplings relevant for oscillation modifications in Eq.~\ref{V2}, let us turn to other potential phenomenological constraints. Despite the small DM masses in this setup, the vector DM produced from inflation is quite ``cold'' and thus does not spoil the successful predictions of standard CDM~\cite{Graham:2015rva}. This makes the introduction of the neutrino coupling potentially dangerous since these relativistic particles can inject energy into the cold DM population and heat it up, or produce a population of hot DM directly.  Although one typically has to worry about the impact of {\it late} kinetic decoupling in ``neutrinophilic DM'' setups (see \cite{Capozzi:2017auw}), here the problem is more severe. In the present case it is required that DM-$\nu$ interactions  {\it never} come into equilibrium. Let's examine the potential problems from this. 

The most dangerous processes that proceeds until late times is (inverse) decays of the sterile, $\nu_{s} \leftrightarrow \nu_{a} + A$. We estimate the rate of this decay in the early Universe to be, $ \Gamma_{\nu_{s} \rightarrow \nu_{a} + A} \simeq \left(\frac{m_{s}}{T}\right)\times \frac{g_A^{2} m_{s}}{8\pi} \times \sin^{2} \theta$. As we approach the temperature $T \simeq m_{s}$ two things happen: (1) the boost factor in front of $ \Gamma_{\nu_{s} \rightarrow \nu_{a} + A}$ disappears and (2) the sterile population is rapidly depleted. Crucially this processes must remain always out of equilibrium ($\Gamma < H$) to avoid producing a sizable abundance of hot DM from the decay products. Notice that $\Gamma_{\rm inv}$ is maximal around $T\simeq m_{s}$. Thus by requiring that $\Gamma < H$ at $T=m_{s}$ we conservatively obtain 
\be 
\label{thermal}
g_A\lesssim \sqrt{\frac{8 \pi m_{s}}{\sin^{2} \theta M_{{\rm Pl}}}} \simeq 6 \times 10^{-13}~\left(\frac{0.1}{\sin \theta}\right)~\sqrt{\frac{m_{s}}{{\rm eV}}}
\ee
for any value of $m_{A}$ kinematically accessible.

At higher $T$ processes like annihilation $\bar{\nu} \nu \leftrightarrow AA$ will dominate since $\Gamma_{ann} \sim g_A^{4} T$. However we find that the bound obtained from such annihilation processes is always less stringent than Eq.~(\ref{thermal}). Notice that is in contrast with what was found in~\cite{Huang:2017egl}. This difference can be understood as follows. First, unlike~\cite{Huang:2017egl}, we are taking the vector $A_{\mu}$ to constitute the DM and not merely to be constrained by contributions to the relativistic energy density around MeV temperatures as in~\cite{Huang:2017egl}. Second since we are operating in the extremely small coupling regime the rate for annihilation processes ($\Gamma_{ann} \propto g_A^{4} $) is extremely suppressed compared to inverse decays ($\Gamma_{inv} \propto g_A^{2}$). Lastly, note that in the $10^{-5}$ eV regime the vector is stable while the sterile neutrino opens an additional path for thermalization, $\nu_{s} \leftrightarrow \nu_{a} + A$.

{Next, consider the possibility that the sterile neutrino is absent at the relevant epochs in the early Universe. This could happen for example in models with heavy sterile neutrinos. Then integrating out the steriles the only impact of the vector is through an effective coupling to active neturinos, $\mathscr{L} \supset g_{\rm eff}~ \bar{\nu}_{a} \gamma_{\mu} \nu_{a} A^{\mu}$. Then $\bar{\nu} + \nu \rightarrow A + A$ annihilation can produce a hot population of $A$'s. Requiring the rate of this annihilation process $\Gamma_{{\rm ann}} \sim g_{{\rm eff}}^{4} T$ to be sub-Hubble at CMB times implies, $g_{{\rm eff}} \lesssim 10^{-7}$. }

%{\color{blue} Is this too weak to include here?}

Now we turn to present-day bounds. A well known constraint on the properties of neutrino interactions comes from the observation of the neutrinos observed in 1987A. The $\bar{\nu}_{e}$ detected from the supernova can be removed by scattering on background DM or neutrinos. We find that the neutrino background places the strongest constraints despite the larger DM density. This comes about because the $t$-channel $\nu$ scattering via $\phi$ exchange is strongly enhanced at small-$t$. Although a small momentum is exchanged, the scattering can ``sterilize'' the neutrino flavor content, $\nu_{a} + \nu_{s} \rightarrow \nu_{s} + \nu_{s}$~\cite{Cherry:2014xra,Cherry:2016jol}.

To compute the optical depth for this process we include the possibility of local gravitational clustering of neutrinos in the Milky Way. This has been estimated to lead to clustered number densities $f_{{\rm cl}}\simeq 20$ above the smooth cosmic background~\cite{Ringwald:2004np}. Then the optical depth is $\tau = \int_{\ell_{0}}^{\ell_{LMC}}\sigma_{\nu}~ n_{\nu}(\ell) ~ d \ell $,
and we model the neutrino density along the line of sight as $n_{\nu}(\ell) = n_{{\rm C\nu B}}~\left( f_{{\rm cl}} (\ell_{0}/\ell)^{2} + 1\right)$, where the first term is the MW contribution and the last term is the cosmic contribution (taking $\propto r^{-2}$ as a model of the MW matter density). Using $\sigma_{\nu} = \sin^{2} \theta g_A^{4}/(4\pi m_{A}^{2})$ we find that requiring the optical depth to be less than unity implies %({\color{blue} Shouldn't this scale as $1/m_s^2$, since $m_s\gg m_\phi$?})
\be 
g_A \lesssim 2 \times 10^{-7}~\left(\frac{m_{A}}{10^{-5}~{\rm eV}}\right)^{1/2}~\left(\frac{0.1}{\sin \theta}\right)^{1/2}
\ee
We note that a similar process can modify the PeV astrophysical flux seen by IceCube. However the extremely small mediator masses produce larger effects in MeV supernova neutrinos. We note that in models where the sterile are heavy enough to be irrelevant at these scales, one can still obtain a SN1987A bound from the energy losses an incoming suffers from significant $t\neq 0$ momentum-transfering collisions~\cite{Kolb:1987qy}.

\section{Effective Hamiltonian: Illustrative examples}
\label{sec:examples}

It is instructive to discuss a few scenarios in which (\ref{effH}) can be studied analytically. We consider either $b=0$ or $a=0$, since combinations of the two parameters always result in subleading effects.

 \subsection{$a_{\mu\tau}\neq0$}

Neglecting $\Delta m^2_{\rm sol}$ and $\theta_{13}$ the SM effective Hamiltonian has no ${e\alpha}$ entries and we are justified to consider the $\mu\tau$ subset of the full 6 by 6 matrix (\ref{effH}):
\ba\label{toy1}
H_{\rm eff}=
\left(
\begin{matrix}
H_i\sigma^i & {\bf 0}_{2\times2} \\
{\bf 0}_{2\times2} & \widetilde{H}_i\sigma^i
\end{matrix}
\right)+{\rm diagonal},
\ea
where $H_i$ are real numbers defined by:
\ba\label{expa}
H_1&=&\frac{\Delta m^2_{\rm atm}}{4E}\sin2\theta_{23}+{\rm Re}\left[a_{\mu\tau}\right]\\\no
H_2&=&{\rm Im}\left[a_{\mu\tau}\right]\\\no
H_3&=&-\frac{\Delta m^2_{\rm atm}}{4E}\cos2\theta_{23},
%\sqrt{H_1^2+H_2^2}&=&\left|\frac{\Delta m^2_{\rm atm}}{4E}\sin2\theta_{23}+a_{\mu\tau}\right|\\\no
%&=&\frac{\Delta m^2_{\rm atm}}{4E}\sin2\theta_{23}+{\rm Re}[a_{\mu\tau}]+\frac{1}{2}\frac{({\rm Im}[a_{\mu\tau}])^2}{\frac{\Delta m^2_{\rm atm}}{4E}\sin2\theta_{23}}+{\cal O}(a^3).
\ea
whereas $\widetilde{H}_i$ is obtained replacing $a\to -a^*$ in ${H}_i$. The oscillation probability is immediately derived from
\ba\label{ab0}
e^{-iH_{\rm eff}L}&=&{\bf 1}_{4\times4}\,{\cos(L/L_{\rm osc})}-iL_{\rm osc}H_{\rm eff}\,{\sin(L/L_{\rm osc})}
%\\\no&=&{\cos(L/L_{\rm osc})}\left(
%\begin{matrix}
%{\bf 1}_{2\times2} & 0 \\
%0 & {\bf 1}_{2\times2}
%\end{matrix}
%\right)-iL_{\rm osc}{\sin(L/L_{\rm osc})}
%\left(
%\begin{matrix}
%H_i\sigma^i & {\bf 0}_{2\times2} \\
%{\bf 0}_{2\times2} & H_i\sigma^i
%\end{matrix}\right)
\\\no
1/L^2_{\rm osc}&\equiv&\frac{1}{4}{\rm tr}[H_{\rm eff}H_{\rm eff}]%={H^{\rm SM}_i{H_i^{\rm SM}}+|b_{\mu\tau}|^2}.
\ea
and reads $P_{\alpha\to\beta}=(H_1^2+H_2^2)L_{\rm osc}^2\,\sin^2(L/L_{\rm osc})$. For intermediate $E$,
\ba\label{Pcanc}
P_{\mu\leftrightarrow\tau}-P_{\mu\leftrightarrow\tau}^{(a=0)}&\propto&{\rm Re}[a_{\mu\tau}]\\\no
&+&\frac{2E}{\Delta m^2_{\rm atm}\sin2\theta_{23}}\left[({\rm Im}[a_{\mu\tau}])^2-(1-2\cos(4\theta_{23}))({\rm Re}[a_{\mu\tau}])^2\right]+{\cal O}(a^3),
\ea 
and a partial cancellation between ${\rm Re}[a_{\mu\tau}], {\rm Im}[a_{\mu\tau}]$ may take place. We see this correlation at play in our numerical results.

 \subsection{$b\neq0$}

An analytic solution cannot be written in a simple form when also the SM effective hamiltonian is included. However, the analysis simplifies considerably at high energy, when we can approximately write
\ba
H_{\rm eff}=
\left(
\begin{matrix}
{\bf 0}_{3\times3} & b \\
b^\dagger & {\bf 0}_{3\times3}
\end{matrix}
\right),
\ea
and one finds
\ba\label{toy2}
e^{-iH_{\rm eff}L}&=&{\bf 1}_{6\times6}-iL_{\rm osc}H_{\rm eff}\,{\sin(L/L_{\rm osc})}-2L^2_{\rm osc}H^2_{\rm eff}\,{\sin(L/2L_{\rm osc})}\\\no
%&=&\left(
%\begin{matrix}
%{\bf 1}_{3\times3} & 0 \\
%0 & {\bf 1}_{3\times3}
%\end{matrix}
%\right)-iL_{\rm osc}\sin({L}/{L_{\rm osc}})\left(
%\begin{matrix}
%0 & b \\
%b^\dagger & 0
%\end{matrix}
%\right)\\\no
%&-&2L_{\rm osc}^2\sin^2({L}/{2L_{\rm osc}})\left(
%\begin{matrix}
%bb^\dagger & 0 \\
%0 & b^\dagger b
%\end{matrix}\right)\\\no
1/L^2_{\rm osc}&\equiv&\frac{1}{4}{\rm tr}[H_{\rm eff}H_{\rm eff}].%{|b_{e\mu}|^2+|b_{e\tau}|^2+|b_{\mu\tau}|^2}={{\rm tr}[b^\dagger b]/2}.
\ea
The flavor selection rules imply that neutrino/anti-neutrino transitions are proportional to $b$ and determine the form of the transition probability, $P(\nu_\alpha\to\bar\nu_\beta)={|b_{\alpha\beta}|^2}L_{\rm osc}^2\sin^2(L/L_{\rm osc})$. Neutrino/neutrino conversions are also possible, but the flavor selection rules now imply that the amplitude is controlled by $bb^\dagger$, so $P(\nu_\alpha\to\nu_{\beta\neq\alpha})=4\sin^4(L/2L_{\rm osc}) |b_{\alpha\gamma}|^2|b_{\beta\gamma}|^2L_{\rm osc}^4$: flavor conversions may happen only if two or more independent terms in $b$ are turned on.

When the SM contribution is introduced, the oscillation length $L_{\rm osc}$ receives an additional contribution and the relative relevance of $b$ diminishes. This can be explicitly checked assuming $b_{\mu\tau}\neq0$, $a=0$, and a constant SM potential for simplicity. Neglecting the solar mass splitting and $\theta_{13}$ the model is a slight modification of (\ref{toy1}):
\ba
H_{\rm eff}=
\left(
\begin{matrix}
H_i\sigma^i & b_{\mu\tau}i\sigma^2 \\
-b_{\mu\tau}^*i\sigma^2 & H_i\sigma^i
\end{matrix}
\right),
\ea
where the real numbers $H_i$ are the same as in~(\ref{expa}), but now with $a=0$. One immediately recovers (\ref{ab0}). As expected, neutrino/neutrino transitions are mediated by the off-diagonal component of the standard potential, $P(\nu_\mu\to\nu_\tau)=\sin^2(L/L_{\rm osc}){(H_1^2+H_2^2)}L_{\rm osc}^2$, whereas $P(\nu_\mu\to\bar\nu_\tau)=\sin^2(L/L_{\rm osc}){|b_{\mu\tau}|^2}L_{\rm osc}^2$. Note that with $|b_{\alpha\gamma}|<|H_{1,2}|$ neutrino/neutrino transitions induced by $b$, being controlled by a higher power of $b$, are more suppressed compared to $\nu_\alpha\leftrightarrow\bar\nu_\beta$. Also, as a consequence of the flavor selection rules there is no special value of $b\neq0$ for which a suppression of the probabilities is attained and expect no particular correlation between ${\rm Im}(b_{\alpha\beta})$ and ${\rm Re}(b_{\alpha\beta})$.

%\vspace{2cm}

\bibliographystyle{JHEP}

\bibliography{nu}
\end{document}